\pdfoutput=1

\documentclass[reprint, sn-nature]{revtex4-2}

\usepackage{datetime2}
\usepackage{natbib}
\usepackage{graphicx}
\usepackage{dcolumn}
\usepackage{bm}
\usepackage{courier}
\usepackage{soul}
\usepackage{amsmath}
\usepackage{amssymb}
\usepackage{xcolor}
\usepackage{soul}
\usepackage[utf8]{inputenc}
\usepackage[pagewise]{lineno}

\begin{document}

\title{Optimizing quantum gates towards the scale of logical qubits}

\author{Paul V. Klimov$^1$}\thanks{corresponding author, pklimov@google.com}
\author{Andreas Bengtsson$^1$}
\author{Chris Quintana$^1$}
\author{Alexandre Bourassa$^1$}
\author{Sabrina Hong$^1$}
\author{Andrew Dunsworth$^1$}

\author{Kevin J. Satzinger$^1$}
\author{William P. Livingston$^1$}
\author{Volodymyr Sivak$^1$}
\author{Murphy Y. Niu$^1$}
\author{Trond I. Andersen$^1$}
\author{Yaxing Zhang$^1$}

\author{Desmond Chik$^1$} 
\author{Zijun Chen$^1$}
\author{Charles Neill$^1$}
\author{Catherine Erickson$^1$}
\author{Alejandro Grajales Dau$^1$}

\author{Anthony Megrant$^1$}
\author{Pedram Roushan$^1$}
\author{Alexander N. Korotkov$^{1,2}$}
\author{Julian Kelly$^1$}
\author{Vadim Smelyanskiy$^1$}
\author{Yu Chen$^1$}
\author{Hartmut Neven$^1$}
\affiliation{$^1$Google Quantum AI}
\affiliation{$^2$Department of Electrical and Computer Engineering, University of California, Riverside, CA}

\DeclareRobustCommand{\hlcyan}[1]{{\sethlcolor{white}\hl{#1}}}

\date{\today}
\maketitle 

\textbf{
A foundational assumption of quantum error correction theory \cite{95qec, 97qec} is that quantum gates can be scaled to large processors without exceeding the error-threshold for fault tolerance \cite{14threshold}. Two major challenges that could become fundamental roadblocks are manufacturing high performance quantum hardware and engineering a control system that can reach its performance limits.  The control challenge of scaling quantum gates from small to large processors without degrading performance often maps to non-convex, high-constraint, and time-dynamic control optimization over an exponentially expanding configuration space. Here we report on a control optimization strategy that can scalably overcome the complexity of such problems. We demonstrate it by choreographing the frequency trajectories of 68 frequency-tunable superconducting qubits to execute single- and two-qubit gates while mitigating computational errors. When combined with a comprehensive model of physical errors across our processor, the strategy suppresses physical error rates by $\sim3.7\times$ compared with the case of no optimization. Furthermore, it is projected to achieve a similar performance advantage on a distance-23 surface code logical qubit with 1057 physical qubits. Our control optimization strategy solves a generic scaling challenge in a way that can be adapted to a variety of quantum operations, algorithms, and computing architectures. 
}

\section{Introduction}

\begin{figure*}[!t]
\includegraphics[width=\textwidth]{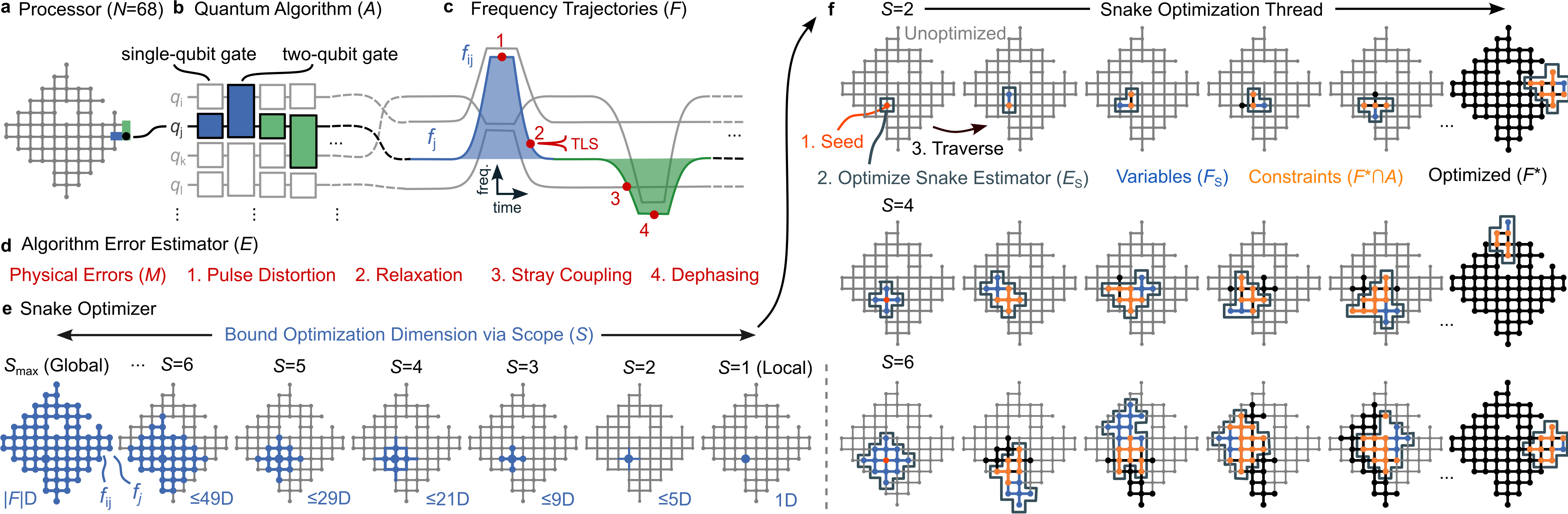}
\caption{\textbf{Frequency optimization}. (a) Our quantum processor with $N=68$ frequency-tunable superconducting transmon qubits represented as a graph. Nodes are qubits (e.g. black dot) and edges are engineered interactions between them (e.g. blue and green lines). (b) A quantum algorithm ($A$) comprising single- and two-qubit gates with one qubit ($q_j$) distinguished. (c) Corresponding qubit frequency trajectories ($F$), parameterized by single-qubit idle ($f_j$ for qubit $q_j$) and two-qubit interaction ($f_{ij}$ for $q_i$ and $q_j$) frequencies. Quantum computational errors depend strongly on frequency trajectories since most physical error mechanisms are frequency dependent (red dots are non-exhaustive examples). Namely, pulse distortion errors (1) happen during large frequency excursions. Relaxation errors (2) happen around relaxation hotspots, for example due to two-level-system defects (TLS, horizontal resonance). Stray coupling errors (3) happen during frequency collisions between coupled computational elements. Dephasing errors (4) happen towards lower frequencies, where qubit flux-sensitivity grows. (d) We leverage our understanding of physical errors to estimate the algorithm's error ($E$) and then optimize it with respect to qubit frequency trajectories. (e) We employ the Snake optimizer, which can solve optimization problems at an arbitrary dimension, controlled by the scope parameter ($S$). These graphs show possible idle (nodes) and interaction (edges) frequency optimization variables (blue) at one Snake optimization step, for scopes ranging from $S=S_{\text{max}}$ (global limit) to $S=1$ (local limit). (f) Snake optimization threads for three scopes (increase downwards). Snake's high configurability enables it to scalably overcome frequency optimization complexity and be adapted to a variety of quantum operations, algorithms, and architectures.}
\label{fig:main_problem_solution}
\end{figure*}

Superconducting quantum processors have demonstrated elements of surface code quantum error correction \cite{22_krinner_surf_walraff, surf17zhao, surf} establishing themselves as promising candidates for fault-tolerant quantum computing. Nonetheless, imperfections in hardware and control introduce physical errors that corrupt quantum information \cite{nielsen00} and could limit scalability. Even if a large enough quantum processor with a high enough performance limit to implement error correction can be manufactured, there is no guarantee that a control strategy will be able to reach that limit.

Frequency-tunable architectures \cite{85martinis, 07koch, dicarlo2009, fluxonium2009, barends2013, 17_dicarlo_fo, heavyflux2018, sup2, sup3, 22rigetti_fo, 2022fluxoniumsiddiqi, 2022alibabaflux, 22_krinner_surf_walraff} are uniquely positioned to mitigate computational errors since most physical error mechanisms are frequency dependent \cite{diracfermigoldenrule, qengguide, peter, decoherence2005, oliverdd, dec2022, dec2023, houckstraycoupling, cryoscope},  (Fig. \ref{fig:main_problem_solution}a-d). However, to leverage this architectural feature, qubit frequency trajectories must be choreographed over quantum algorithms to simultaneously execute quantum operations while mitigating errors. 

Choreographing frequency trajectories is a complex optimization problem due to engineered and parasitic interactions among computational elements \cite{houckstraycoupling} and their environment \cite{purcell_spontaneous_1946, mullertls, oliverboxmodes2021}, hardware \cite{2021_laseranneal} and control \cite{cryoscope} inhomogeneities, performance fluctuations \cite{klimtls}, and competition between error mechanisms. Mathematically, the problem is non-convex, highly constrained, time-dynamic, and expands exponentially with processor size. 

Past research into overcoming these complexities employed frequency partitioning strategies \cite{Barends2014, fopatent} that either faced difficulties scaling with realistic hardware imperfections or whose scalability is not well understood \cite{17_dicarlo_fo, fo1, 22rigetti_fo, 22_krinner_surf_walraff}. \hlcyan{To overcome the limitations of these strategies, we proposed the ``Snake" optimizer \mbox{\cite{snake}} and employed an early version in past reports \mbox{\cite{sup1, qaoa21, mcewencosmic, miaoleakage, scrambling, topological, timecrystal, majorana, anyon, sup4}}. However, an optimization strategy has not been developed around it, it has not been rigorously benchmarked, and large enough processors to investigate its scalability have only recently become available}. Whether high performance configurations exist at scale and whether they can be quickly discovered and stabilized are open questions.

Here we address these questions by developing a control optimization strategy around Snake that can scalably overcome the complexity of problems like frequency optimization within the high performance, high stability, and low runtime requirements of an industrial system. \hlcyan{The strategy introduces generic frameworks for building processor-scale optimization models, training them for various quantum algorithms, and adapting to their unique optimization landscapes via Snake. This flexible approach can be applied to a variety of quantum operations, algorithms, and architectures. We believe it will be an important element in scaling quantum control and realizing commercially valuable quantum computations. 

We investigate the prospects of this strategy for optimizing quantum gates for error correction in superconducting qubits}. We demonstrate that it strongly suppresses physical error rates, approaching the surface code threshold for fault tolerance on our processor with tens of qubits. \hlcyan{To pave the way towards much larger processors, we demonstrate Snake “healing” and “stitching”, which were designed to stabilize performance over long timescales and geometrically parallelize optimization.  Finally, we introduce a simulation environment that emulates our quantum computing stack and combine it with optimization, healing, and stitching to project the scalability of our strategy towards thousands of qubits.}

\section{Quantum hardware}
Our hardware platform is a Sycamore processor with $N=68$ frequency-tunable transmon qubits on a two-dimensional lattice. Engineered tunable coupling exists between $109$ nearest-neighbors \cite{charlesthesis, feiyan, surf}. We configure the processor \hlcyan{to execute the surface code gate set}, which includes single-qubit XY rotations (SQ) and two-qubit controlled-Z (CZ) gates \cite{dicarlo2009}. SQ gates are implemented via microwave pulses resonant with qubits’ respective $|0\rangle \leftrightarrow |1\rangle$ \textit{idle frequencies} ($f_i$ for qubit $q_i$ executing SQ$_i$). CZ gates are implemented by sweeping neighboring qubits into $|11\rangle \leftrightarrow |02\rangle$ resonance near respective \textit{interaction frequencies} ($f_{ij}$ for qubits $q_i$ and $q_{j}$ executing CZ$_{ij}$) and actuating their couplers.  The $N=68$ idle and $\sim2N=109$ interaction frequencies - which constitute one frequency configuration $F$ with dimension $\sim 3N = |F| = 177$ - parameterize qubit frequency trajectories, which we seek to optimize.

\section{Performance benchmark}
We evaluate the performance of frequency configurations via the parallel two-qubit cross-entropy benchmarking algorithm (CZXEB) \cite{boixo, sup1}. CZXEB executes cycles of parallel SQ gates followed by parallel CZ gates (see SI), benchmarking them in a context representative of many quantum algorithms. \hlcyan{Most relevant to this study is that CZXEB reflects the structure of the surface code's parity checks and has empirically served as a valuable performance proxy of logical error \mbox{\cite{surf}}}. The processed output of CZXEB is the \textit{benchmark distribution} $e_{c}$, in which each value is one qubit pair's average error per cycle $e_{c, ij}$, which includes error contributions from respective $\text{SQ}_i$, $\text{SQ}_j$, and $\text{CZ}_{ij}$ gates. Benchmarks are generally not normally distributed across a processor and are thus reported via percentiles as $50.0\%^{(97.5 - 50)\%}_{(2.5 - 50)\%}$ and plotted as quantile boxplots. The wide range from $2.5\%$ to $97.5\%$ is the distribution spread, which spans $\pm 2\sigma$ standard deviations for normally distributed data.

\section{Optimization model}
We approach frequency optimization as a model based problem. In turn, we must define an \textit{algorithm error estimator} $E$ that is representative of the performance of the target quantum algorithm $A$ at the optimizable frequency configuration $F$ (Fig. \ref{fig:main_problem_solution}d). This problem is hard because the estimator must be fast for scalability, predictive for scaling projections, \hlcyan{and physical for metrology investigations}. We introduce a flexible framework for overcoming these competing requirements that can be adapted to define the optimization landscapes a variety of quantum operations, algorithms, and architectures (see SI). 

Our framework corresponds to the decomposition $E({F}|{A},{D})=\sum_{g\in A}\sum_{m\in M}w_{g,m}({A})\epsilon_{g,m}({F}_{g,m}|D)$. The sums are over all gates $g\in A$ and known physical error mechanisms $m\in M$. $\epsilon_{g,m}$ are algorithm-independent error components that depend on some subset of frequencies $F_{g,m}\subseteq F$ and can be computed from relevant characterization data $D$. $w_{g,m}$ are algorithm-dependent weights that capture algorithmic context via training on benchmarks that are sufficiently representative of $A$. Defining the estimator thus maps to defining the target quantum algorithm, the algorithm-independent error components, and then training the algorithm-dependent weights.

We set our target quantum algorithm to \hlcyan{CZXEB to gear the estimator towards the surface code's parity checks}. Furthermore, since CZXEB is also our benchmarking algorithm, we can associate the performance of optimized frequency configurations with our optimization strategy. We then define error components corresponding to dephasing \mbox{\cite{peter, decoherence2005, oliverdd}}, relaxation \mbox{\cite{decoherence2005, dec2022, dec2023}}, stray coupling \mbox{\cite{houckstraycoupling}}, and frequency-pulse distortion \mbox{\cite{cryoscope}} \hlcyan{over qubit frequency trajectories}. The relevant characterization data include qubit flux-sensitivity spectra, energy-relaxation rate spectra, parasitic stray coupling parameters, and pulse distortion parameters, which are measured prior to optimization. Finally, we train the weights \hlcyan{via a protocol that we developed specifically to reduce the risk of overfitting \mbox{\cite{iterativelearningklimov}} (see SI). It constrains weights via homogeneity and symmetry assumptions and then leverages the frequency tunability of our architecture to train them} on single- and two-qubit gate benchmarks taken in configurations of variable complexity. 

The resulting algorithm error estimator represents a comprehensive understanding of physical errors in our processor. It spans $\sim 4 \times 10^{4}$ error components, only has $16$ trained weights \hlcyan{for the full processor}, and is trained and tested on $\sim 6500$ benchmarks. Despite its scale, it can still be evaluated $\sim100\text{ times}/\text{second}$ on a desktop. Furthermore, it can predict CZXEB cycle errors in the wide range $\sim 3 - 40 \times 10^{-3}$ within two factors of experimental uncertainty (see SI). \hlcyan{In total, the estimator fulfills our speed, predictivity, and physicality requirements}. 

\section{Optimization strategy}

Finding an optimized frequency configuration from the algorithm error estimator maps to solving ${F}^* = \text{argmin}_{F} E$. This problem is hard for several reasons. First, all $|F|\sim3N$ idle and interaction frequencies are interdependent due to engineered and parasitic interactions between nearest and next-nearest neighbor qubits. Second, the estimator has numerous local minima since most error mechanisms and hardware constraints compete, and since it is built from noisy characterization data. Finally, there are $\sim k^{|F|}\sim k^{3N}$ possible configurations, where $k$ is the number of options per frequency, as constrained by hardware and control specifications and inhomogeneities. In total, the problem is highly-constrained, non-convex, and expands exponentially with processor size. We developed the Snake optimizer  \cite{snake} to scalably overcome the complexity of control optimization problems like frequency optimization. 

Snake implements a graph based algorithm that maps the variable frequency configuration ${F}$ onto a graph and then launches an optimization thread from some seed frequency (Fig. \ref{fig:main_problem_solution}e-f). It then finds all unoptimized frequencies ${F}_{S}$ within a neighborhood whose size is bounded by the scope parameter $S$, and constructs the \textit{Snake estimator} $E_{S}$. $E_{S}$ contains all terms in $E$ that depend only on ${F}_{S}$, which serve as optimization variables, and previously optimized frequencies $F^*$ that are algorithmically relevant $F^*\cap A$, which serve as fixed constraints. Snake then solves $F^*_S=\text{argmin}_{F_{S}} E_S$, updates ${F}^*$, traverses, and repeats until all frequencies have been optimized. Frequency configurations are typically optimized from multiple seeds in parallel and the one that minimizes the algorithm error estimator is benchmarked.

Snake's favorable scaling properties are derived from the \textit{scope} $S$, which tunes the greediness of its optimization between the local and global limits \cite{snake}. By tuning the scope within $1\leq S \leq S_{\text{max}}$, we can bound the number of frequencies optimized at each traversal step to $1\leq|{F}_S|\leq|F|$, where $|F_S|\sim S^2$ and $S_{\text{max}} \sim \sqrt{3N}$ in two dimensions (Fig. \ref{fig:main_problem_solution}e). In turn, we can split one complex $\sim 3N$-dimensional problem over $\sim k^{3N}$ configurations into $\sim 3N / S^2$ simpler $\sim S^2$-dimensional problems over $\sim k^{S^2}$ configurations each. Such splitting terminates at $S=1$, where Snake optimizes $\sim 3N$ 1-dimensional problems over $\sim k^{1}$ configurations each. Importantly, the intermediate dimensional problems with $S<S_{\text{max}}$ are exponentially smaller than the global problem and independent of processor size. 

Snake is not expected to discover globally optimal configurations. However, if it can find sufficiently performant configurations \hlcyan{for the target quantum algorithm - for example with errors below the fault-tolerance threshold \mbox{\cite{14threshold}}} - it will solve the scaling complexity problem. Namely, we will not be faced with an exponentially expanding problem as our processors scale, but linearly more problems with bounded configuration spaces. \hlcyan{Furthermore, since Snake's seed strategy, traversal strategy, inner-loop optimizer, and scope are highly configurable, it should be adaptable to overcome similar scaling complexities in other control problems and hardware (see SI)}.

\hlcyan{}

\section{Validating performance}
\begin{figure}[t!]
\includegraphics[width=\linewidth]{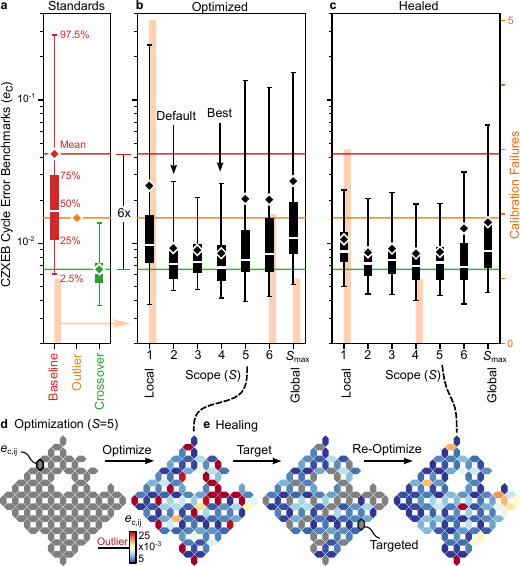}
\caption{\textbf{Optimization and healing performance}.
(a) CZXEB cycle error benchmarks ($e_c$, boxes, left axis) and calibration failures (orange bars, right axis in (c)) for the random baseline (red), outlier (orange diamond), and crossover (green) performance standards used to evaluate frequency configurations and our optimization strategy. Each box shows the 2.5, 25, 50, 75, and 97.5th percentiles and mean (see annotations on the baseline). The standards' means are extended across panels for comparison. (b) Benchmarks for configurations optimized at different scopes ($S$) ranging from $S=1$ (local limit) to $S=S_{\text{max}}$ (global limit). Intermediate dimensional optimization ($2\leq S \leq 4$) \hlcyan{outperforms both local and global optimization}, finding configurations near the crossover standard. \hlcyan{$S=4$ performs best but $S=2$ offers a better balance between performance and runtime (see SI) and is set as our default.} (c) Benchmarks for each configuration in (b) after healing, which significantly suppresses performance outliers. In (a), (b) and (c), each box corresponds to one configuration. (d) Benchmark heatmaps illustrating optimization and (e) healing of targeted gates in the $S=5$ configuration. Each hexagon corresponds to the cycle error for one pair ($e_{c,ij}$). Performant gates are blue, outliers are red, and unoptimized and targeted gates are grey.}
\label{fig:main_scope_experiment}
\end{figure}

To experimentally investigate whether Snake can actually find performant frequency configurations at some intermediate dimension, we optimize our processor at scopes ranging from $S=1$ (177 $1D$ local problems) to $S=S_{\text{max}}$ (one $177D$ global problem) and benchmark CZXEB. We evaluate configurations by comparing their benchmarks against three performance standards (Fig. \ref{fig:main_scope_experiment}a). First, the \textit{baseline standard} references benchmarks taken in a random frequency configuration, which establishes the average performance of the hardware and control system without frequency optimization $e_c=16.7^{+267.1}_{-10.6}\times10^{-3}$ at $N=68$). Second, the \textit{outlier standard} references a constant cycle error, above which gates are considered performance outliers ($e_c=15.0\times10^{-3}$ for all $N$). Third, the \textit{crossover standard} references published benchmarks from the same processor that reached the surface code's \textit{crossover regime}, which approaches the error correction threshold ($e_c=6.2^{+7.6}_{-2.5}\times10^{-3}$ at $N=49$) \cite{surf, 14threshold}. This standard establishes what we consider high performance, while recognizing that much higher performance will be necessary to implement error correction in practice.  

The wide performance gap between the baseline and crossover standards is closed via frequency optimization (Fig. \ref{fig:main_scope_experiment}b). Namely, intermediate dimensions ($2\leq S \leq 4$) approach the crossover standard ($e_c=7.2^{+19.9}_{-2.5}\times10^{-3}$ in $\sim 130$ seconds at $S=2$) while suppressing performance outliers, with $<10\%$ of gates above the outlier standard and $<0.5\%$ failing calibrations, which prevent benchmarking. However, local ($e_c=9.8^{+231.8}_{-6.0}\times10^{-3}$ in $\sim 6$ seconds at $S=1$) and global ($e_c=10.8^{+145.0}_{-5.7}\times10^{-3}$ in $\sim 6,500$ seconds at $S=S_{\text{max}}$) optimization only marginally outperform the baseline standard. The optimal scope is $S=4$ ($\leq21$D), but we default to $S=2$ ($\leq5$D), which offers a better balance between performance and runtime (see SI). Next we interpret. 

\begin{figure*}[t!]
\includegraphics[width=\textwidth]{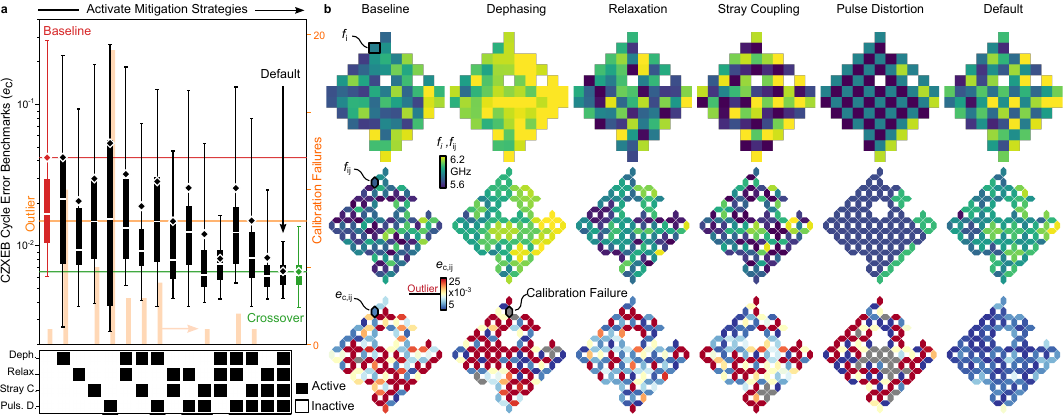}
\caption{\textbf{Optimization performance versus error mitigation strategy.} (a) CZXEB cycle error benchmarks ($e_{c}$, black) and calibration failures (orange bars, rightmost axis) for configurations optimized with all combinations of dephasing, relaxation, stray coupling, and frequency-pulse distortion error mitigation strategies activated (see lower matrix). The random baseline (red), outlier (orange), and crossover (green) standards are shown for comparison. (b) (upper) Idle frequency ($f_i$), (center) interaction frequency ($f_{ij}$), and (lower) cycle error ($e_{c,ij}$) heatmaps for (leftmost) the baseline standard with no mitigation strategies activated, (central) configurations with only one strategy activated, and (rightmost) the default configuration with all strategies activated. As more mitigation strategies are progressively activated (from left to right in (a)), cycle errors and calibration failures trend downwards, highlighting the importance of metrology on the performance of our optimization strategy. 
}
\label{fig:main_mitigation_experiment}
\end{figure*}

First, the fact that we see performance variations between configurations illustrates that poor frequency choices cannot be compensated for by other components of our control system (see SI) and that optimization is critical. Second, the fact that local optimization underperforms illustrates that frequency optimization is a non-local problem and that tradeoffs between gates must be considered. Third, the fact that global optimization underperforms even after an hour of searching illustrates the difficulty of navigating the configuration space even on our relatively small processor. Finally, the fact that relatively low intermediate dimensions found the most performant configurations is consistent with relatively local engineered and parasitic interactions and suggests that Snake can navigate our architecture's configuration space in a way that should scale to larger processors. 

\section{Stabilizing performance}

Stabilizing performant configurations is as difficult and important as finding them. Namely, a processor's optimization landscape constantly evolves and performance outliers emerge on timescales ranging from seconds to months, with the most catastrophic due to TLS defects fluctuating into the path of qubit frequency trajectories \cite{klimtls}.  Unfortunately, even a low percentage of outliers can significantly degrade the performance of a quantum algorithm \cite{surf}. However, re-optimizing all gates of a processor when a low percentage of outliers are detected is unscalable from a runtime perspective and introduces the risk of degrading performant gates.

By design, \textit{Snake healing} can surgically re-optimize outliers, nominally much faster than full re-optimization, and without degrading performant gates \cite{snake}. To investigate the viability of healing, we heal all configurations generated by the variable-scope experiment described above, targetting gates as described in the SI (Fig. \ref{fig:main_scope_experiment}d-e). From the perspective of stability, the progressively worse configurations emulate the performance of our processor over progressively longer timescales following optimization. Healing suppresses outliers by $\sim48\%$ averaged over configurations, typically runs $>10\times$ faster than full reoptimization, and rarely degrades performant gates. Furthermore, heals can be applied repetitively and parallelized for sufficiently sparse outliers. These results demonstrate the viability of healing for scalably suppressing outliers to stabilize performance. 

\section{Impact of Metrology}

We now consider the impact of the algorithm error estimator's composition on Snake's performance. In particular, the dephasing, relaxation, stray coupling, and pulse distortion error components may be interpreted as distinct error mitigation strategies that can be activated independently. To isolate their impact and to understand their interplay, we progressively activate them in all combinations, optimize, and benchmark CZXEB (Fig. \mbox{\ref{fig:main_mitigation_experiment}a)}. 

To build intuition for the impact of each error mitigation strategy, we inspect frequency configurations optimized with only one mitigation strategy activated (Fig. \ref{fig:main_mitigation_experiment}b). Most are visually structured, with \hlcyan{inhomogeneities arising from fabrication imperfections in the processor's parameters}. Dephasing mitigation biases qubits towards their maximum frequencies, where flux sensitivity vanishes \mbox{\cite{07koch}}. Relaxation mitigation biases qubits away from relaxation hotspots driven by \hlcyan{coupling to the control \mbox{\cite{07koch}} and readout circuitry \mbox{\cite{purcell_spontaneous_1946, evanreadout}}, packaging environment \mbox{\cite{oliverboxmodes2021}}}, and random TLS defects \mbox{\cite{mullertls}}. Stray-coupling mitigation disperses qubits to avoid frequency collisions between parasitically coupled gates. \hlcyan{Finally, pulse-distortion mitigation biases idles towards a multi-layered checkerboard, with neighbors at one of two symmetric $|11\rangle \leftrightarrow |02\rangle$ CZ resonances \mbox{\cite{dicarlo2009}}, and interactions towards resonance between the idles, to minimize frequency excursions. The inversion of frequencies at the eastern edges of the processor was triggered by fabrication imperfections that broke the symmetry between CZ resonances. This observation highlights non-trivial interplay between error mitigation and hardware inhomogeneities.}

Interestingly, while some of these mitigation strategies alone may find performant configurations at the scale of several qubits, none of them substantially outperform the random baseline configuration at the scale of our processor. As we progressively activate mitigation strategies, competition between error mechanisms causes frequency configurations lose visual structure, while performance approaches the crossover standard. \hlcyan{Analyzing error contributions in optimized configurations, we confirm that activating mitigation strategies selectively and effectively suppresses their corresponding error components, while only weakly impacting others (see SI). These results support our interpretation of error components as error mitigation strategies and that our optimizer can effectively reconcile their competition and suppress them}. More generally, they highlight the importance of error metrology on the performance of our optimization strategy.

\section{Performance scalability}

\begin{figure}[t!]
\includegraphics[width=\linewidth]{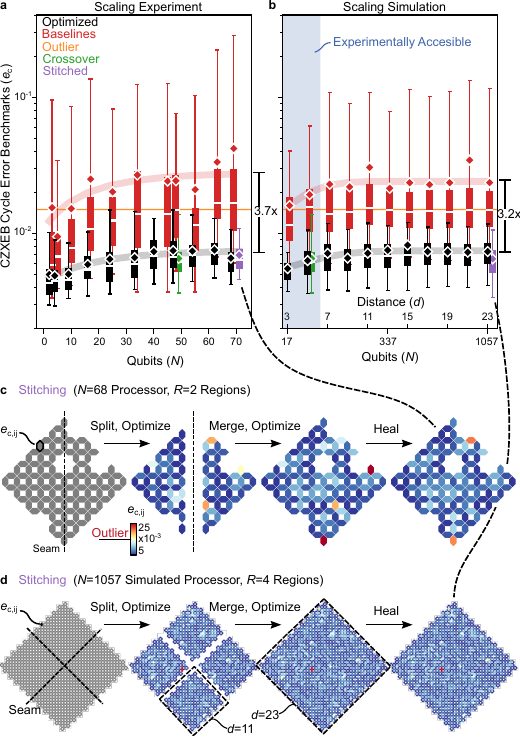}
\caption{\textbf{Optimization scalability}. (a) Experimental and (b) simulated CZXEB cycle error benchmarks ($e_{c}$) in optimized (black) and unoptimized baseline (red) configurations of variable size. Simulated processors have size and connectivity corresponding to surface code logical qubits with distance $d$. The crossover standard (green), outlier standard (orange), and stitched configurations (purple) are shown for comparison. The solid lines are fits of the saturation model to the optimized (black) and baseline (red) benchmark means. Some boxes have been horizontally shifted to reduce overlap. In (a), $N<40$ boxes combine benchmarks from multiple configurations to boost statistics. In (b), the x axis is linear in $d$ with $N = 2d^2 - 1$ and the shaded region illustrates the experimentally accessible regime of our processor. (c) Benchmark heatmaps illustrating stitching of our $N=68$ processor and (d) $N=1057$ simulated processor. Outliers are not substantially amplified at seams, which is our primary concern. We note that stitching the $d=23$ logical qubit with $R=4$ is equivalent to stitching four $d=11$ logical qubits.}
\label{fig:main_scaling_experiment}
\end{figure}

We are finally ready to investigate Snake's scalability. To do so, we conduct a scaling experiment that may be valuable for evaluating the prospects of any quantum hardware and control system. Namely, we optimize, heal, and benchmark hundreds of configurations of our processor ranging in size from $N=2$ to $68$ (Fig. \ref{fig:main_scaling_experiment}a). As before, we reference the crossover standard. However, we now reference multiple baseline standards that correspond to unoptimized random configurations of variable size. Despite the irregular shapes of some configurations, we find surprisingly clear scaling trends. 

CZXEB benchmarks grow and then saturate in both optimized and unoptimized configurations. Furthermore, mean cycle errors are well-represented by the model $\langle e_{c}(N)\rangle= e_{sat} - e_{scale} \exp(-N / N_{sat})$, where $N_{sat}$ is the qubit saturation constant, $e_{scale}$ is the error penalty in scaling gates from small to large systems, and $e_{sat}$ is the saturated error. Fitting this model to the empirical benchmarks, we find that optimized configurations saturate near the crossover standard, with best-fit parameters $N_{sat}=22\pm10$ $(\pm 1\sigma)$, $e_{scale}=3.1\pm0.4\times10^{-3}$, and $e_{sat}=7.5\pm 0.4\times10^{-3}$.

To estimate Snake's performance advantage, we make several comparisons. From the empirical benchmarks, we compare the mean cycle errors $\langle e_{c}^{base}\rangle / \langle e_{c}^{snake}\rangle$ in isolation ($N=2$) and in parallel at scale ($N=68$), which are $3.1 \pm 0.5$ and $6.4 \pm 1.0$, respectively. Remarkably, the optimized $N=68$ configuration outperforms unoptimized $N=2$ configurations by $2.3 \pm 0.4\times$. Finally, the optimized $N=68$ configuration has a $\sim 40\times$ narrower benchmark distribution spread. From the saturation model, we compare the scaling penalty $e^{base}_{scale}/e^{snake}_{scale}$ and the saturated cycle errors $e^{base}_{sat}/e^{snake}_{sat}$, which are $5.6\pm1.8$ and $3.7\pm0.7$, respectively. These comparisons illustrate that Snake achieves a significant performance advantage to $N=68$. 

To investigate Snake’s future scalability, we simulate much larger processors than manufactured to date. To do so, we developed a generative model that can generate simulated processors of arbitrary size and connectivity with simulated characterization data that are nearly indistinguishable from our processor \cite{gmpatent} (see SI). We generate simulated processors ranging in size from $N=17$ to $1057$, with connectivity corresponding to distance-3 to 23 surface code logical qubits ($d=3$ to $23$ with $N = 2d^2 - 1$) \cite{fowler2019low}.  We optimize simulated processors exactly like our processor and predict CZXEB benchmarks via our estimator. Simulated benchmarks reproduce the saturation trends seen in experiment, building trust in our simulation environment and results (Fig. \ref{fig:main_scaling_experiment}b). Furthermore, they project that Snake’s performance advantage should scale to a $d=23$ logical qubit with $N=1057$. 

\section{Runtime scalability}
Despite the promising performance outlook, practically scaling to thousands of qubits will require Snake to be geometrically parallelized. Namely, even though optimization runtimes scale nearly linearly with processor size ($\sim3.6\pm0.1$ seconds added per qubit at $S=2$), $N=1057$ threads take $\sim1.4$ hours. This exceeds our runtime budget of $0.5$ hours, \hlcyan{which was chosen for compatibility with operating large surface codes (see SI)}. 

By design, \textit{Snake stitching} can split a processor into $R$ disjoint regions, optimize them in parallel, and stitch configurations \cite{snake}. Stitching leads optimization runtimes to scale sub-linearly with processor size, which should enable scalability towards $N\sim10^4$ with $R=128$ within our runtime budget in principle. In practice, however, stitching risks amplifying outliers at seams, where Snake must reconcile constraints between independently optimized configurations.

To investigate the viability of stitching, we stitch and heal our $N=68$ processor with $R=2$ (Fig. \ref{fig:main_scaling_experiment}c) as well as an $N=1057$ ($d=23$) simulated processor with $R=4$ (Fig. \ref{fig:main_scaling_experiment}d). \hlcyan{We chose convenient stitch geometries, but believe they will ultimately need to be optimized (see SI)}. Experimental data are limited, but outliers are not amplified at seams and stitched configurations perform as well as their unstitched counterparts ($e_c=6.4^{+4.4}_{-1.8}\times10^{-3}$ for $N=68$ and $e_c=6.3^{+4.3}_{-2.9}\times10^{-3}$ for $N=1057$). Finally, we note that stitching the $d=23$ logical qubit with $R=4$ is equivalent to stitching four $d=11$ logical qubits into a 4-logical-qubit processor \cite{fowler2019low}, which illustrates how larger surface codes may be optimized.

\section{Outlook}

\hlcyan{We introduced a control optimization strategy that combines generic frameworks for building, training, and navigating the optimization landscapes presented by a variety of quantum operations, algorithms, and architectures}. It offers a significant performance advantage for quantum gates on our superconducting quantum processor with tens of qubits, approaching the surface code threshold for fault tolerance, and shows promise for scalability towards logical qubits with thousands of qubits. \hlcyan{A recent demonstration of error suppression in a scaled up surface code logical qubit \mbox{\cite{surf}} enabled by this strategy underscores its potential.} 

\hlcyan{Elements of our strategy have also been employed to optimize quantum operations including measurement \mbox{\cite{rosnake}} and SWAP gates \mbox{\cite{qaoa21}}, and quantum algorithms for optimization \mbox{\cite{qaoa21}}, metrology \mbox{\cite{mcewencosmic, miaoleakage}}, simulation \mbox{\cite{scrambling, topological, timecrystal, majorana, anyon}}, and beyond classical computation \mbox{\cite{sup1, sup4}}. The strategy should also find value in quantum hardware beyond superconducting circuits \mbox{\cite{qengguide}}, which face control challenges with similar scaling complexities. Choreographing the trajectories of electrons in quantum dots \mbox{\cite{qdshuttling, qdcontrol, siliconqe, guidosemispins}}, shuttled ions in ion traps \mbox{\cite{ionautoshuffle, ionshufflecompiler, ioncontrol}}, or neutral atoms in reconfigurable atom arrays \mbox{\cite{atomoptimization, atomoptimization2023}} are promising applications (see SI) that are of significant contemporary interest.}

\hlcyan{Looking towards commercially valuable quantum computations}, significant challenges remain. The larger and more performant processors that are necessary to implement them will be susceptible to error mechanisms that are currently irrelevant or yet to be discovered. Furthermore, we expect that stabilizing performance over long computations that may span days \cite{gidney2021} will present significant hurdles. Towards that end, Snake's model-based approach can leverage historical characterization data to forecast and optimize around failures before they happen \cite{forecastingpatent, mltrading}. Finally, even though we expect that model-based optimization will remain critical for injecting metrological discoveries into control optimization \hlcyan{for the foreseeable future}, Snake can also deploy model-free reinforcement learning agents \cite{sutton, 21qctrlrl, sivakrl}, which may reduce the burden of developing performance estimators (\hlcyan{see SI}). The techniques presented here should complement the numerous other control, hardware, and algorithm advancements necessary to implement quantum error correction. 

\section{\label{sec:Contributions}Contributions}
P.V.K. conceived, prototyped, and led the development of the Snake optimizer, algorithm error estimator, and simulated processor generative modeling frameworks. An.B. and C.Q. contributed to engineering the optimizer. Al.B. and A.D. contributed to engineering the generative model. C.Q, Al.B., A.D., K.J.S., M.Y.N., W.P.L., V.S., T.I.A., and Y.Z. contributed to error metrology research. S.H. led the development of parallel calibration infrastructure with engineering contributions from An.B and Z.C.. Finally, A.M., P.R., A.N.K., J.K, V.S., Y.C., and H.N. supported research and development.

\section{\label{sec:Acknowledgements}Acknowledgements} 
We thank the broader Google Quantum AI team for fabricating the processor, building and maintaining the cryogenic system, and general hardware and software infrastructure that enabled this experiment. We also thank Austin Fowler, Alexis Morvan, and Xiao Mi for feedback on the manuscript.

\bibliography{main.bib}
\end{document}


\title{Supplementary information \\Optimizing quantum gates towards the scale of logical qubits}

\author{Paul V. Klimov$^1$}\thanks{corresponding author, pklimov@google.com}
\author{Andreas Bengtsson$^1$}
\author{Chris Quintana$^1$}
\author{Alexandre Bourassa$^1$}
\author{Sabrina Hong$^1$}
\author{Andrew Dunsworth$^1$}

\author{Kevin J. Satzinger$^1$}
\author{William P. Livingston$^1$}
\author{Volodymyr Sivak$^1$}
\author{Murphy Y. Niu$^1$}
\author{Trond I. Andersen$^1$}
\author{Yaxing Zhang$^1$}

\author{Desmond Chik$^1$} 
\author{Zijun Chen$^1$}
\author{Charles Neill$^1$}
\author{Catherine Erickson$^1$}
\author{Alejandro Grajales Dau$^1$}

\author{Anthony Megrant$^1$}
\author{Pedram Roushan$^1$}
\author{Alexander N. Korotkov$^{1,2}$}
\author{Julian Kelly$^1$}
\author{Vadim Smelyanskiy$^1$}
\author{Yu Chen$^1$}
\author{Hartmut Neven$^1$}
\affiliation{$^1$Google Quantum AI}
\affiliation{$^2$Department of Electrical and Computer Engineering, University of California, Riverside, CA}

\DeclareRobustCommand{\hlcyan}[1]{{\sethlcolor{white}\hl{#1}}}

\date{\today}

\maketitle

\tableofcontents


\section{Control System}
\begin{figure}[t!]
\includegraphics[width=\linewidth]{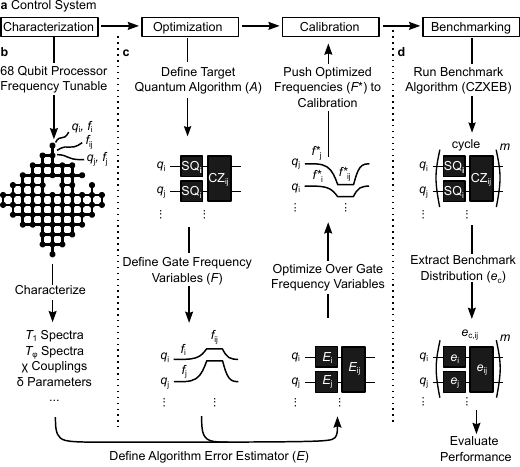}
\caption{
\textbf{Control system}. 
Overviews of our (a) control system and it's (b) characterization, (c) optimization, and (d) benchmarking components, which are detailed in the text.}
\label{fig:si_control_system}
\end{figure}

Here we overview our control system to offer context for where our optimization system exists within our quantum computing stack (Figure \ref{fig:si_control_system}). The following sections describe most of these components in more detail. Our control system learns the control-electronics signals needed to execute quantum gates for some target quantum algorithm ($A$). It can be split into the characterization, optimization, calibration, and benchmarking systems. The characterization system measures data ($D$) that are believed to be necessary to operate and estimate the performance of quantum gates over qubits' operable frequency ranges, including qubits' flux sensitivities versus frequency ($\sim T_{\phi}^{-1}$ spectra), energy-relaxation rates versus frequency ($T_{1}^{-1}$ spectra), parasitic stray coupling parameters ($\chi$ coupling), and frequency-pulse distortion parameters ($\delta$ parameters). The optimization system then estimates the quantum algorithm’s error ($E$) around that data and optimizes it over gate frequencies ($F$). The calibration system then learns the control parameters necessary to execute gates at the optimized frequency configuration ($F^*$) (details in \cite{sup1, surf, optimus}). Finally, the benchmarking system evaluates the performance of the configuration via one or more benchmarking algorithms. If calibration failures were not encountered and if benchmarks exceed some standard of high performance, the quantum algorithm is executed. 

\section{Characterization System}
The characterization system interrogates our hardware's computational elements - including qubits, couplers, and readout resonators - and the control electronics (details in \cite{surf, sup1}). The output is the characterization data $D$, which includes:

\begin{itemize}
\item \textbf{Gate trajectory parameters}
\subitem SQ gate length: $t_{\text{SQ}}=25$ ns.
\subitem CZ gate length: $t_{\text{CZ}}=34$ ns.
\subitem $\mathcal{F}$ State-dependent CZ frequency trajectory \cite{foxen}.
\item \textbf{Hardware and control parameters}
\subitem$T_1$ relaxation time versus frequency (i.e. spectra), from which we estimate relaxation errors.
\subitem $\frac{df}{d\phi}$ flux sensitivity spectra, from which we estimate $\sim T_\phi$ dephasing errors. 
\subitem $\chi$ parasitic stray coupling parameters, from which we estimate stray coupling errors due to parasitic coupling between nearest- and next-nearest-neighbor qubits.
\subitem $\delta$ frequency-pulse distortion parameters, from which estimate errors due to large frequency excursions during CZ gates.
\end{itemize}

The characterization data are used to construct optimization hard bounds and the algorithm error estimator's error components, as discussed below.

\section{The Algorithm Error Estimator}
\hlcyan{The algorithm error estimator can be represented by a variety of models (e.g. a linear model, a neural network, or a quantum simulation). The only strict requirement for optimization is that the estimator be representative of the performance of the target quantum algorithm. For this study, we append several requirements - physicality, speed, and accuracy. These are competing interests that are difficult to satisfy simultaneously. In the following sections we describe how we construct an estimator that can satisfy them.}

\subsection{Construction}
We define the \textit{algorithm error estimator} $E({F}|{A},{D})$ by making several simplifying assumptions:

\begin{enumerate}
\item A quantum algorithm's error can be decomposed into a sum over the \textit{gate error estimators} $E_{g}(F_g|A,D)$ of its constituent gates $g \in A$ \hlcyan{(i.e. CZXEB with 2 qubits and $m$ cycles comprises $2m$ SQ and $m$ CZ gate error estimators}). $F_g$ is the subset of $F$ that are relevant to $E_g$. \hlcyan{This assumption is motivated by the digital error model, which states that gate errors can be added when they are small and not correlated in space or time \mbox{\cite{boixo}}, and which was validated in the context of random circuits \mbox{\cite{sup1}}.}

\item A gate's error can be decomposed into a sum over algorithm-independent \textit{error components} $\epsilon_{g,m}(F_{g, m}|D)$, each of which corresponds to a distinct physical error mechanism $m \in M$. $F_{g, m}$ is the subset of $F_{g}$ that are relevant to estimating $\epsilon_{g,m}$. \hlcyan{This assumption should apply in the limit of small and uncorrelated error components. Error components should be defined for all known error mechanisms, including those that go beyond the assumptions of the digital error model (e.g. stray coupling errors, which are correlated in space). Furthermore, they should be defined for all possible algorithmic contexts (e.g. stray coupling components for all possible combinations of parasitically coupled gates executed concurrently).}

\item The effect of implementing an arbitrary gate within an arbitrary quantum algorithm can be fully encompassed by algorithm-dependent weights $w_{g, m}(A)$. \hlcyan{One unique weight is assigned to, and multiplies, each error component.} The weights filter error components by algorithmic relevance (e.g. weights corresponding to stray coupling errors are only non-zero for concurrent gates) and relative contribution (e.g. weights corresponding to dephasing errors are lower in algorithms with dynamical decoupling than without).

\item The weights can be trained on benchmarks that are sufficiently representative of the target quantum algorithm. \hlcyan{Empirical training injects algorithmic context and can compensate for some of the simplifying assumptions made above and inaccuracies in the error components.}
\end{enumerate}

These assumptions lead to the decomposition:

\begin{align}
\begin{split}
E({F}|{A},{D})=\sum_{g\in A}E_{g}({F}_{g}|{A},{D}) = \\ 
    \sum_{g\in A}\sum_{m \in M}w_{g,m}({A})\epsilon_{g,m}({F}_{g,m}|D)
\end{split}
\end{align}

\hlcyan{This decomposition establishes a powerful and broadly applicable framework:}

\begin{itemize}
    \item \hlcyan{Compatible with physicality, speed, and accuracy.}
    \item \hlcyan{Error components can be added as new error mechanisms are discovered.}
    \item \hlcyan{Error components can be interchanged to transfer the estimator between hardware architectures (see Section \mbox{\ref{otherhardware}}).}
    \item \hlcyan{Weights can be re-trained to transfer the estimator between quantum algorithms.}
    \item \hlcyan{Optimization variables ($F$ here) can be interchanged to adapt the estimator to other control variables, provided that the error components are defined in terms of those variables.}
    \item \hlcyan{Equivalent to the basis-expansion method in machine learning \mbox{\cite{elements, norvig, introstatlearning}}, bridging the domains and facilitating knowledge transfer.}
\end{itemize}

The assumption that errors can be added in this way is expected to break down in certain limits - particularly in highly structured quantum circuits. \hlcyan{Fortunately, methods like randomized compiling \mbox{\cite{rcemerson, rcsiddiqi, rcqec}} offer the potential to recompile those algorithms into compliance}. Finally, even when our assumptions cannot be fulfilled, we expect that our algorithm error estimator can still serve as a valuable optimization proxy.

\subsection{Error components}

We define error components for the dephasing \cite{peter, decoherence2005, oliverdd}, relaxation \cite{decoherence2005, murphytls, dec2022, dec2023}, stray coupling \cite{houckstraycoupling}, and pulse distortion \cite{cryoscope, Foxen_2019} error mechanisms. We leverage textbook physics, published literature, and extensive metrology of our quantum and classical control hardware. Furthermore, since runtime is a critical scaling consideration, we engineer error components for speed via software optimizations, physical simplifications, and mathematical approximations, some of which trade against accuracy. 

Here we provide \hlcyan{key} error components $\epsilon_{g,m}$ for gates $g$ and mechanisms $m$, the relevant frequencies $F_{g,m}$ and characterization data $D$, \hlcyan{and how they scale with processor size $N$. We note that each error mechanism is segmented into multiple error components as described in detail below. When referencing a mechanism, we consider all such components together}.

\begin{enumerate}
\item{Single qubit gate error components $g = \text{SQ}_i$}

\begin{enumerate}
\item $m =$ Dephasing error on SQ$_i$ due to $q_i$.
\subitem $\epsilon_{g, m}(F_{g, m}) \propto t_{\text{SQ}} / T_{\phi, i}(f_i) \propto t_{\text{SQ}} \frac{df_i}{d\phi_i}$.
\subitem $F_{g, m} = \{f_i\}$
\subitem $D =$ Flux-sensitivity spectrum for $q_i$, $\frac{df_i}{d\phi_i}$ and SQ$_i$ length, $t_{\text{SQ}}$.
\subitem \hlcyan{Scale: $N$ (1 per SQ gate)}

\item $m =$ Relaxation error on SQ$_i$ due to $q_i$.
\subitem $\epsilon_{g, m}(F_{g, m}) \propto t_{\text{SQ}} / T_{1, i}(F_{g, m})$.
\subitem $F_{g, m} = \{f_i\}$
\subitem $D =$ Relaxation spectrum for $q_i$, $T_{1, i}(F_{g, m})$ and SQ$_i$ length, $t_{\text{SQ}}$.
\subitem \hlcyan{Scale: $N$ (1 per SQ gate)}

\item $m =$ Stray coupling error on SQ$_i$ due to SQ$_j$.
\subitem $\epsilon_{g, m}(F_{g, m}) \propto \text{Lorentzian}(F_{g, m}; \chi_{ij})$.
\subitem $F_{g, m} = \{f_i, f_j\}$
\subitem $D =$ Stray coupling parameters between SQ$_i$ and SQ$_j$, $\chi_{ij}$.
\begin{equation}
\text{\hlcyan{Scale: }} 32N
    \begin{cases}
      N \text{ SQ gates}\\
      8 \text{ parasitic qubits each}\\
      4 \text{ collisions each}
    \end{cases}       
\end{equation}

\end{enumerate}

\item{Two qubit gate error components $g = \text{CZ}_{ij}$}.
\begin{enumerate}

\item $m =$ Dephasing error on CZ$_{ij}$ due to $q_i$ undergoing some state-dependent frequency trajectory $\mathcal{F}_i(F_{g, m})$.
\subitem $\epsilon_{g, m}(F_{g, m}) \propto \int_{\mathcal{F}_i(F_{g, m})}  \frac{df}{d\phi} \frac{df}{dt}^{-1} df$ 
\subitem $F_{g, m} = \{f_i, f_j, f_{ij}\}$
\subitem $D =$ Flux-sensitivity spectrum $\frac{df}{d\phi}$ for $q_i$ and CZ$_{ij}$ frequency trajectory $\mathcal{F}_i$.
\begin{equation}
\text{\hlcyan{Scale:}} \sim 8N
    \begin{cases}
      \sim2N\text{ CZ gates} \\
      4 \text{ input states each}
    \end{cases}       
\end{equation}

\item $m =$ Relaxation error on CZ$_{ij}$ due to $q_i$ undergoing some state-dependent frequency trajectory $\mathcal{F}_i(F_{g, m})$.
\subitem $\epsilon_{g, m}(F_{g, m}) \propto \int_{\mathcal{F}_i(F_{g, m})} T_{1, i}^{-1}(f) \frac{df}{dt}^{-1} df$ 
\subitem $F_{g, m} = \{f_i, f_j, f_{ij}\}$
\subitem $D =$ Relaxation spectrum $T_{1, i}(f_i)$ for $q_i$ and CZ$_{ij}$ frequency trajectory $\mathcal{F}_i$.
\begin{equation}
\text{\hlcyan{Scale:}} \sim 8N
    \begin{cases}
      \sim2N \text{ CZ gates}\\
      4 \text{ input states each}
    \end{cases}       
\end{equation}

\item $m =$ Stray coupling error on CZ$_{ij}$ due to CZ$_{kl}$.
\subitem $\epsilon_{g, m}(F_{g, m}) \propto \text{Lorentzian}(F_{g, m}; \chi_{ij,kl})$.
\subitem $F_{g, m} = \{f_i, f_j, f_k, f_l, f_{ij}, f_{kl}\}$
\subitem $D =$ Stray coupling parameters between CZ$_{ij}$ and CZ$_{kl}$, $\chi_{ij,kl}$.
\begin{equation}
\text{\hlcyan{Scale:}} \sim 896N
    \begin{cases}
      \sim2N \text{ CZ gates}\\
      14 \text{ parasitic qubits each}\\
      \sim32 \text{ collisions each}
    \end{cases}       
\end{equation}

\item $m =$ Frequency-pulse distortion error on CZ$_{ij}$ due to $q_i$'s excursion from $f_i$ to $\sim f_{ij}$.
\subitem $F_{g, m} = \{f_i, f_{ij}\}$
\subitem $D =$ Frequency-pulse distortion parameters for CZ$_{ij}$, $\delta_{ij}$.
\subitem \hlcyan{Scale: $\sim2N$ (1 per CZ gate)}

\end{enumerate}
\end{enumerate}

In total, \hlcyan{an $N$ qubit processor has $\sim10^3 N$ error components}. Our $N=68$ processor has $\sim4\times10^{4}$ error components. This number is large due to the high granularity with which we segment error components: 

\begin{itemize}
\item \hlcyan{Each error mechanism is generally segmented into single- and two-qubit error components (e.g. 1a and 2a above for dephasing)}. 

\item \hlcyan{Two-qubit error components typically integrate errors over frequency trajectories $\mathcal{F}$ according to the hardware implementation and the local temporal order of gates within the quantum algorithm. For CZXEB, they may integrate over a trapezoidal trajectory that links idles and interactions}.

\item Each error component is segmented by gate. For example, we consider 68 ($N$) components for SQ dephasing (1a above) \hlcyan{and 109 ($\sim 2N$) components for CZ pulse distortion (2d above)}.

\item \hlcyan{Some error components are segmented by input state. For example, we consider the 4 input states $|00\rangle, |01\rangle, |10\rangle, |11\rangle$ for CZ dephasing and relaxation (2a and 2b above).}

\item Stray coupling error components \hlcyan{are segmented by the number of parasitically coupled qubits and the number of frequency collisions each (1c and 2c above). For example, we consider 8 parasitic qubits for each SQ gate (nearest and next-nearest neighbors) and 14 parasitic qubits for each CZ gate (nearest and next-nearest neighbors, excluding the CZ qubits)}. Furthermore, we consider 4 collisions between each pair of parasitically coupled SQ gates (all possible collisions between the qubits' respective $|0\rangle \leftrightarrow |1\rangle$ and $|1\rangle \leftrightarrow |2\rangle$ transition frequencies) \hlcyan{and 32 collisions between each pair of parasitically coupled qubits executing CZ gates.}
\end{itemize}

\vfill\null

\subsection{Runtime}
\begin{figure}[t!]
\includegraphics[width=\linewidth]{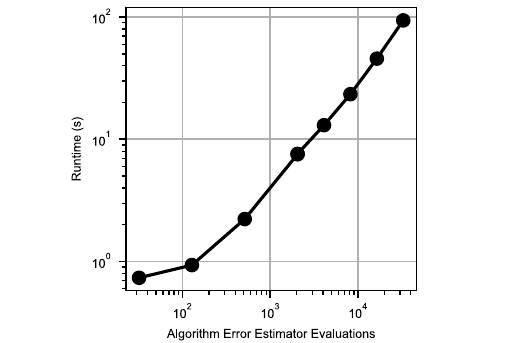}
\caption{\textbf{Algorithm error estimator runtime}. 
Runtime versus the number evaluations of our algorithm error estimator on a high-performance desktop. The estimator is defined over 177 gate frequency variables and comprises $\sim 40,000$ error components. Despite its large scale, it can still be evaluated $>100 \times$/second.}
\label{fig:si_alg_estimator_runtime}
\end{figure}

The majority of optimization runtime is spent evaluating error components. To estimate how quickly we can evaluate our error components, we consider evaluation runtimes for our $N=68$ processor's 177 gate-frequency variable algorithm error estimator on a high-performance desktop (Figure \ref{fig:si_alg_estimator_runtime}). We can evaluate the estimator $>100 / \text{second}$, which translates into $\lesssim1 / \text{ second}$ for evaluating all error components of a gate on average.

\subsection{\hlcyan{Extensions to other hardware}}
\label{otherhardware}
\hlcyan{Extending the algorithm error estimator to control optimization problems in other hardware will require defining new error components. While the connection between error components in our transmon qubits \mbox{\cite{07koch}} and other superconducting qubits \mbox{\cite{qengguide}} is reasonably clear, the connection to entirely different hardware is not. To illustrate the versatility of our framework, we connect our error components to control objectives when choreographing the spatial trajectories of reconfigurable atom arrays \mbox{\cite{atomoptimization, atomoptimization2023}}. The objective of minimizing atom loss and heating by avoiding crossing spatial trajectories is analogous to our stray coupling components, which penalize crossing frequency trajectories. The objective of minimizing atom move distance is analogous to our pulse distortion components, which penalize large frequency excursions. The objective of minimizing atoms' vertical extent is similar to our dephasing components, which squeeze frequencies towards their maxima. Similar connections exist to control problems in other hardware, for example shuttling electrons in quantum dots \mbox{\cite{qdshuttling, qdcontrol, siliconqe, guidosemispins}} or shuttling ions in ion traps \mbox{\cite{ionautoshuffle, ionshufflecompiler, ioncontrol, ionshuttling}}.} 

\section{Training the Estimator}

\hlcyan{When training the estimator, we must ensure that it generalizes to unseen configurations of our processor. In turn, we take significant care in mitigating the risk of overfitting \mbox{\cite{elements, norvig, mohri, introstatlearning}}. In traditional statistics and machine learning applications, overfitting is often combated by reducing the complexity of a model by discarding or constraining correlated features - e.g. via principal component analysis - or by suppressing them during training - e.g. via regularization. These approaches are most compatible with models that don't necessarily require interpretability, such as neural networks. However, they are incompatible with our requirement that the estimator be physical. Namely, we must keep all physically relevant error components, even though some of them may be correlated in some scenarios. Next we formalize the overfitting problem and then develop training-data sampling and model-training protocols to overcome it.}

\subsection{Bounding model capacity}
Model capacity is an important concept in machine learning that measures the expressivity of a model and is often associated with the number of trainable parameters \cite{mohri}. It is desirable to select a model for training that has a high enough capacity to be able to represent complex patterns but low enough capacity to reduce the risk of overfitting and the amount of training data necessary. 

For the algorithm error estimator, we equate the model capacity with the number of independent trainable weights. Within this definition, the capacity of our $N=68$ algorithm error estimator is $\sim4\times10^{4}$, since each error component generally has an independent trainable weight. Furthermore, given local engineered and parasitic coupling, we expect the capacity to scale linearly with processor size. 

Given the considerations mentioned above, taking all weights to be independent trainable parameters is neither practical nor scalable. To bound the capacity of our estimator, we make several reasonable assumptions:

\begin{itemize}
\item \textbf{Homogeneity}: The parameters of our processor are homogeneous enough such that weights for the same error components for the same gate types are equal. For example 1(a) above, this assumption leads to $w_{g, m} = w_{g', m}$ for $g=$SQ$_i$ and $g'=$SQ$_j$. 
\item \textbf{Symmetry}: The weights of some error components must be equal due to symmetry. For example 2(a) above, this assumption leads to $w_{g, m} = w_{g, m'}$, where $m$ and $m'$ correspond to the respective trajectories $\mathcal{F}_i$ for the input states $|0, 1\rangle$ and $|1, 0\rangle$.
\end{itemize}

By applying these assumptions, we reduce our algorithm error estimator's capacity from $\sim 4\times10^4$ to $16$. Furthermore, capacity does not grow with processor size. We thus believe this is a scalable strategy for bounding the capacity of our algorithm error estimator.

\subsection{Sampling training data}
Building a sufficiently large and diverse training dataset such that the trained estimator generalizes to complex and unseen scenarios is a complex problem in benchmarking and statistics \cite{elements, introstatlearning}. Towards that end, we leverage the flexibility of our quantum processor architecture to accomplish the following: 

\begin{itemize}
\item \textbf{Multiple training targets}: We train on both SQRB and CZXEB benchmarks. 
\item \textbf{Isolate error components}: We employ the frequency-tunability of our architecture to benchmark gates in configurations of variable complexity. For example, to isolate the relaxation and dephasing error components, we benchmark gates in sparse configurations with negligible stray coupling. As another example, to controllably introduce stray coupling, we benchmark gates in denser configurations with only nearest- or next-nearest neighbor gates benchmarked simultaneously. 
\item \textbf{Boost statistical leverage}: We employ the frequency-tunability of our architecture to vary the relative strength of each error component. For example, we benchmark gates near and far from their respective qubits' maximum frequencies to generate data with high leverage in dephasing. As another example, we benchmark gates near and far from two-level-system (TLS) defects to generate data with high leverage in relaxation.
\end{itemize}

For this study, we sampled $\sim6500$ benchmarks, averaging $> 100$ training benchmarks per independent trainable weight, over several weeks. Although time consuming, the trained estimator remains viable over months, and is trivial to refine as benchmarks accumulate over time. Furthermore, the runtime cost of acquiring training data in parallel scales inversely with processor size (i.e. since the number of gates that can be benchmarked in parallel scales as $\sim N$), which is especially favorable from a scalability perspective. Finally, we believe that if a sufficiently complete understanding of a processor and the control system is available, the training data may be augmented via generative modeling methods similar to those used to generate our simulated processors \cite{gmpatent}. 

\subsection{Training}
\begin{figure}[t!]
\includegraphics[width=\linewidth]{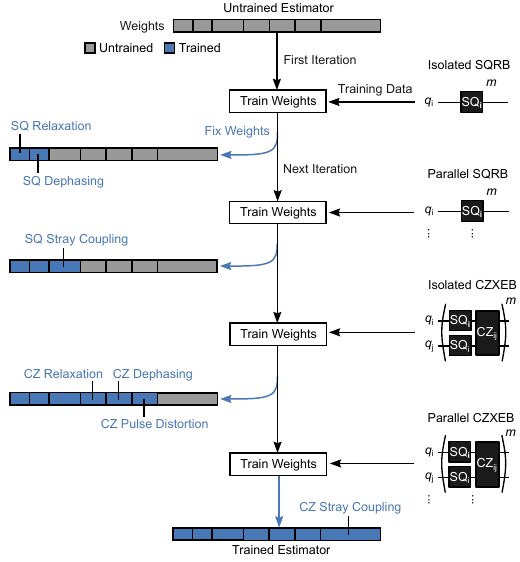}
\caption{\textbf{Iterative training protocol}. We train the algorithm error estimator over multiple iterations on benchmarks of increasing complexity. \hlcyan{At the start of the protocol (top), the estimator's weights are untrained (grey). The different segments separated by vertical lines correspond to the weights for distinct subsets of error components. At each iteration, distinct subsets of error components and their weights are isolated and trained on training data corresponding to judiciously chosen benchmarks (right)}. Namely, isolated benchmarks isolate relaxation, dephasing, and pulse distortion error components, while parallel benchmarks introduce stray coupling error components. \hlcyan{Once trained, the weights are fixed (left) to their trained values (blue) and the training protocol proceeds onto the next iteration. At the end of the protocol (bottom), all of the estimator's weights have been trained}.}
\label{fig:si_training_protocol}
\end{figure}

\begin{figure*}[p!]
\includegraphics[width=\textwidth]{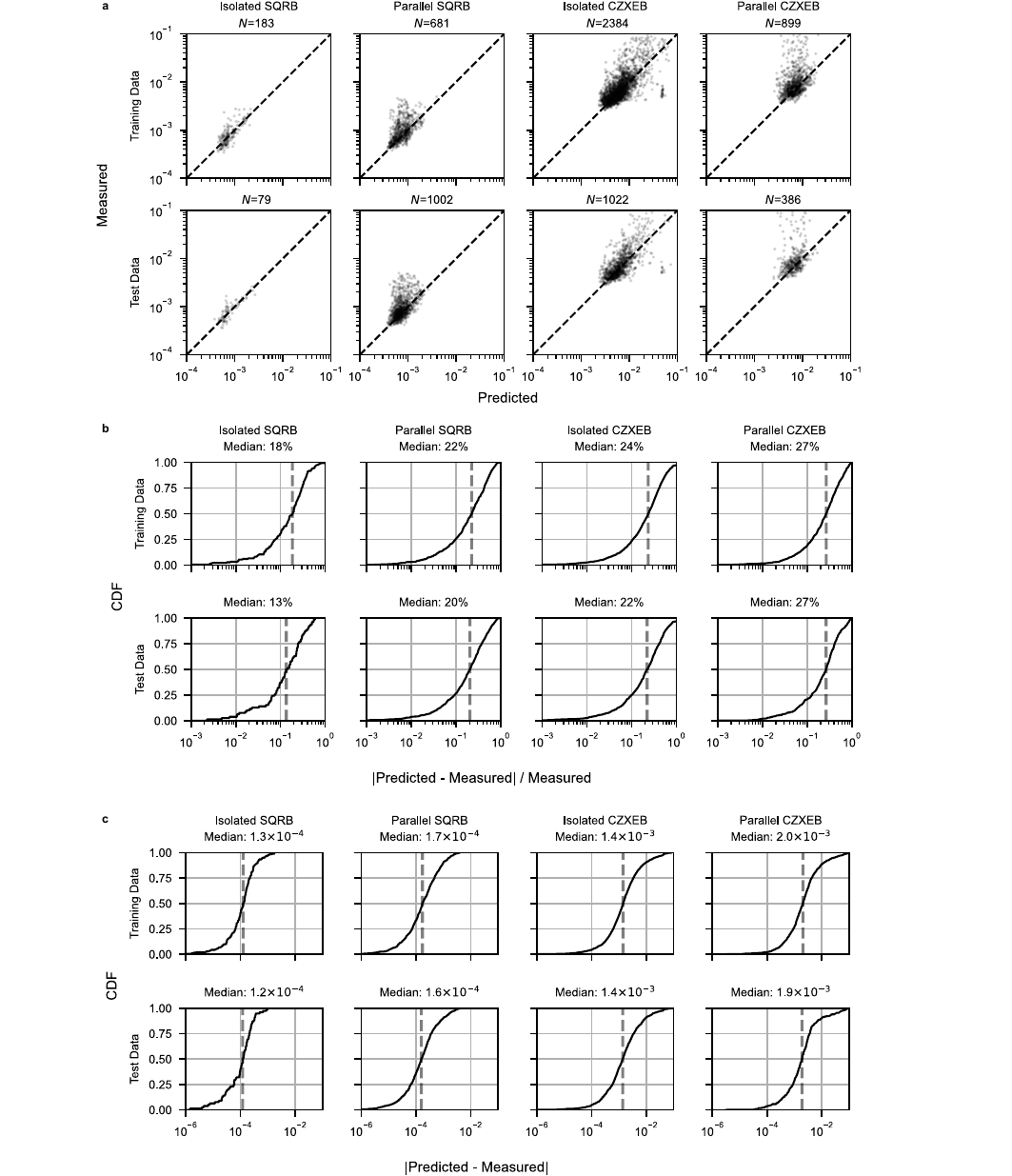}
\caption{\textbf{Algorithm error estimator predictions versus measurements}.
(a) Measured versus predicted SQRB and CZXEB benchmarks for the (upper) training and (lower) test data and for Isolated and Parallel benchmarks. $N$ is the number of samples. The data were taken in frequency configurations of variable complexity. Deviations from the diagonal dashed line - where predictions match measurements - correspond to inaccuracies. (b) Cumulative distribution functions (CDF) of the Inaccuracy and (c) Relative Inaccuracy for the data in (a). The dashed vertical lines in (b) and (c) correspond to the distribution medians.}
\label{fig:si_validation}
\end{figure*}

To train the algorithm error estimator's weights, we developed an \textit{iterative supervised learning} protocol \cite{iterativelearningklimov}. It progressively trains and constrains distinct subsets of the estimator's weights over multiple training iterations (Figure \ref{fig:si_training_protocol}). Early iterations train the estimator on isolated benchmarks, which isolate the weights corresponding to relaxation and dephasing. Later iterations train the estimator on parallel benchmarks, which introduce weights corresponding to stray coupling. We note that this protocol was designed specifically to further suppress the capacity of the model trained at any given iteration.

We implement training iterations within the standard machine learning framework \cite{elements, introstatlearning, norvig, tensorflow}, using $\sim 60\%$ of the available benchmarks for training and the remaining $\sim 40\%$ for computing accuracy metrics (below). The error components are the training features and the error benchmarks are the training targets. At each iteration, we train using the Adam optimizer \cite{adam} with a mean-absolute-error cost function, which is less susceptible to outliers than the more standard mean-squared-error. Since the trained weights are not generally useful, they will not be presented. 

\subsection{Accuracy}

\begin{figure}[t!]
\includegraphics[width=\linewidth]{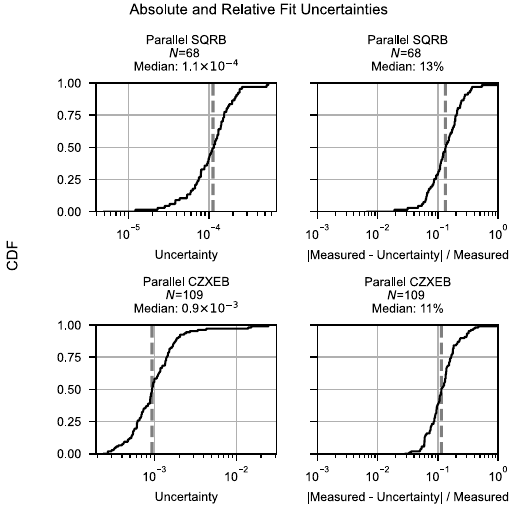}
\caption{\textbf{Experimental uncertainty}.
(left) Experimental uncertainty and (right) relative experimental uncertainty for (upper) Parallel SQRB and (lower) Parallel CZXEB benchmarks for one 68 qubit configuration. The dashed vertical lines correspond to the medians. $N$ is the number of samples.}
\label{fig:si_exp_uncertainty}
\end{figure}

We evaluate the accuracy of our trained estimator on the test data, which was not used during training. Our trained estimator can predict both SQRB and CZXEB benchmarks in arbitrary configurations. To evaluate the accuracy of these predictions, we consider the following metrics and quote the medians (Figure \ref{fig:si_validation}): 

\begin{itemize}
    \item \textbf{$\text{Inaccuracy} = |\text{Predicted} - \text{Measured}|$}
    \subitem Isolated SQRB $\sim 1.2 \times 10^{-4}$
    \subitem Parallel SQRB $\sim 1.6 \times 10^{-4}$
    \subitem Isolated CZXEB $\sim 1.4 \times 10^{-3}$
    \subitem Parallel CZXEB $\sim 1.9 \times 10^{-3}$
    \item \textbf{$\text{Relative Inacc.} = \text{Inaccuracy} / \text{Measured}$}
    \subitem Isolated SQRB $\sim 13 \%$
    \subitem Parallel SQRB $\sim 20 \%$
    \subitem Isolated CZXEB $\sim 22 \%$
    \subitem Parallel CZXEB $\sim 27 \%$
\end{itemize}

\begin{figure*}[pth!]
\centering
\vspace*{1cm}
\includegraphics[width=\textwidth]{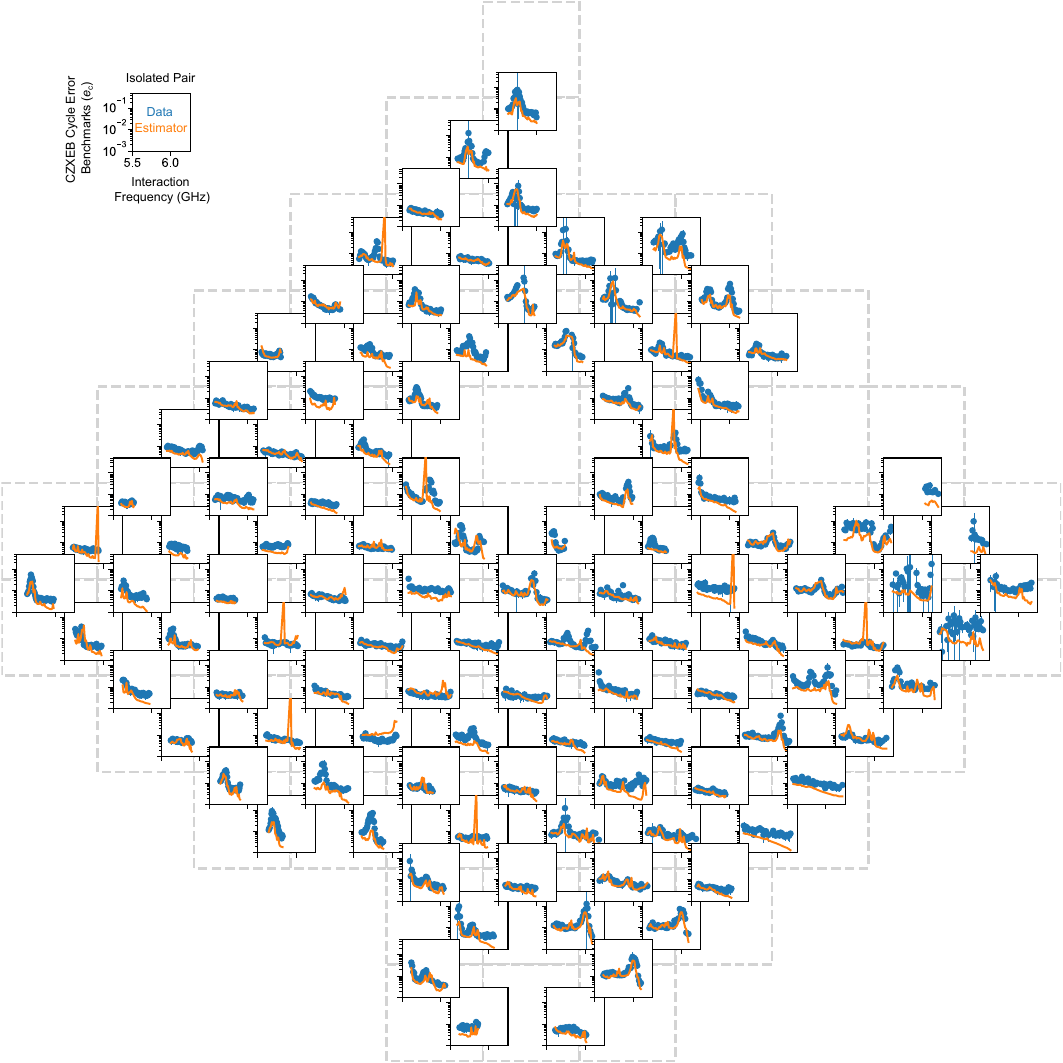}
\caption{\textbf{Isolated CZXEB benchmarks versus interaction frequency, with idles fixed}. These data can be interpreted as a linecut of the processors much higher dimensional error landscape, where all idles and interactions are variables. The inset shows the common scale, with the data (blue) and algorithm error estimator predictions (orange) overlaid. \hlcyan{Error bars correspond to 68\% confidence intervals and are typically smaller than the data points}. The error axis (vertical) is logarithmically spaced to highlight small inaccuracies. The grey boxes represent the qubits of our processor and map to the processor graph in Figure \ref{fig:si_control_system}b. The algorithm error estimator can reproduce complex error patterns, some of which span an order of magnitude. However, there are also clear inaccuracies, some of which are discussed in the text.}
\label{fig:si_xeb_vs_freq}
\end{figure*}

Although the medians are a concise statistic, they oversimplify the situation. Namely, inaccuracy metrics typically depend on and increase with the magnitude of benchmark values. These dependencies suggest that defining a \textit{trust region}, within which we trust our estimator's predictions, may be more useful. 

We define the trust region to be the range of Isolated and Parallel CZXEB test data over which the estimators inaccuracy is simultaneously $\lesssim 1/2\times$ the measured error and $\lesssim 2\times$ the experimental uncertainty (Figure \ref{fig:si_exp_uncertainty}). We interpret these as signal-to-noise metrics. Finally, we note that CZXEB benchmarks are the hardest to predict, requiring accurate SQ and CZ gate error predictions.

\begin{itemize}
\item \textbf{Trust region}:  $\sim3$ to $\sim40\times10^{-3}$
\end{itemize}

Therefore, we can trust our estimator within a wide range that spans one order of magnitude. 

\subsection{Innacuracy sources}
We expect our estimator's predictions to deviate from measurements for reasons including: 

\begin{itemize}
    \item Simplifications and/or approximations made to the error components to suppress runtimes.
    \item Less training data towards higher errors. 
    \item Undiscovered physical error mechanisms. 
    \item Control and hardware parameter inhomogeneities.
\end{itemize}

For further insights, we first inspect predicted versus measured data in Figure \ref{fig:si_validation}. The largest inaccuracies are seen towards high parallel CZXEB errors ($\sim 5\%$ of the training and test data). Since similar inaccuracies are not as prevalent in other benchmarks, we believe the primary culprits are: 

\begin{itemize}
    \item Inaccuracies in the CZ stray coupling error components, which are known to be complex. 
    \item The impact of leakage, which can be driven via stray coupling, on CZXEB is not well understood.
\end{itemize}

Second, we inspect the isolated CZXEB data in Figure \ref{fig:si_xeb_vs_freq}. Here the interaction frequencies are swept while the idles are fixed. These data should be interpreted as line cuts of the processor's much higher dimensional error landscape, where all idles and interactions are variables. Isolated CZXEB mostly isolates dephasing and relaxation errors, the latter of which dominate in our system and exhibit the most complex patterns. These data suggest that our estimator can reproduce complex error patterns, some of which span an order of magnitude and exceed the trust region defined above. However, we also observe clear inaccuracies, the most significant of which we believe are due to: 

\begin{itemize}
    \item TLS fluctuations \cite{klimtls} between characterization and benchmarking. 
    \item Underrepresented physics corresponding to qubits interacting with strongly-coupled TLS \cite{murphytls}. 
\end{itemize}

\section{Optimization System}

Our optimization system is based on the Snake optimizer, \hlcyan{which we proposed \mbox{\cite{snakepatent, snake}} as a platform for deploying custom optimization strategies within the demands of an industrial control system. Snake leverages concepts in dynamic programming and graph optimization to offer several key functionalities, which to the best of our knowledge are not offered by any other optimizer:}

\begin{itemize}
    \item \hlcyan{\textbf{Flexibility}: Can implement a wide array of optimization strategies via judicious selection of several parameters (see Section \mbox{\ref{sec:snakeparams}}). Most notably, it can deploy virtually any inner loop optimizer at any dimension between the local and global optimization limits. This customizability facilitates adapting Snake to a variety of optimization landscapes.}
    \item \hlcyan{\textbf{Scalability}: Runtime scales linearly versus processor size without parallelization and sub-linearly with geometric parallelization (``stitching").}
    \item \hlcyan{\textbf{Stability}: Can locally re-optimize performance outliers (``healing") as they emerge over time to stabilize processors over long timescales (i.e. months). Healing is much faster than optimization (see Section \mbox{\ref{sec:runtime_scalability}}) and scales sub-linearly with the number of outliers when parallelized.}
\end{itemize}

\hlcyan{In our past proposal, we established Snake's theoretical foundations and software abstractions. In this section, we map frequency optimization into Snake. We then provide broadly applicable implementations of its parameters and employ them to explore tens of optimization strategies between the local and global optimization limits over thousands of simulated optimization threads. Finally, we promote the most promising strategies to the experiments in the main text.}

\subsection{Optimization variables}
The qubit frequency trajectories used to implement gates can be parameterized by many optimization variables. We allocate one variable per gate: 

\begin{itemize}
\item \textbf{Idle frequency ($f_i$)}: $q_i$'s $|0\rangle \leftrightarrow |1\rangle$ transition frequency where it executes SQ$_i$ \cite{surf}. 
\item \textbf{Interaction frequency ($f_{ij}$)}: $q_i$ and $q_j$'s average uncoupled $|0\rangle \leftrightarrow |1\rangle$ transition frequencies, where they execute CZ$_{ij}$ \cite{surf}.
\end{itemize}

We choose these variables because we have empirically verified that they have a significant impact on gate performance. Although we could add more trajectory variables, for example gate lengths, each such variable exponentially expands the configuration space and presents a scalability barrier. Instead we either fix or locally optimize those variables within our calibration system \cite{sup1}. 

\subsection{Optimization bounds}

Each frequency variable is subject to hard bounds set by the quantum and classical control hardware. The bounds are computed from the characterization data and embed a detailed understanding of our hardware and control systems. Furthermore, they are contracted as much as reasonably possible via physics intuition to reduce the configuration space and thus runtimes. 

Here we list several hard bounds: 

\begin{itemize}
\item \textbf{Idle frequency bounds} 
\subitem Maximum detuning from qubits' maximum frequencies to limit dephasing. 
\subitem Minimum detuning from qubits' readout resonators to limit Purcell relaxation and resonator-induced dephasing \cite{evanreadout, 07koch}. 
\subitem Minimum and maximum detuning from the LO to limit microwave pulse-distortion due to the finite DAC bandwidth. 
\item \textbf{Interaction frequency bounds}
\subitem Maximum detuning from qubits' maximum frequencies to limit dephasing. 
\subitem Minimum detuning from qubits' readout resonators to limit Purcell relaxation and resonator-induced dephasing \cite{evanreadout, 07koch}. 
\end{itemize}

For $N=68$ configurations, these hard bounds constrain SQ and CZ gates to average operating bandwidths $\sim450 \pm 120$ ($\pm 1\sigma$) MHz and $\sim635 \pm 112$ MHz, respectively. \hlcyan{From these operating bandwidths, together with a 2 MHz hardware discretization, we can estimate the average number of idle and interaction frequency options ($k$ in Section V of the main text) and the total number of frequency configurations for the processor}:

\begin{itemize}
    \item \hlcyan{\textbf{Idle frequency options}: $\sim225 \pm 60$.}
    \item \hlcyan{\textbf{Interaction frequency options}: $\sim318 \pm 56$.}
    \item \textbf{Frequency configurations}: $\sim 2^{1437}$ total configurations \hlcyan{for our processor}. This number significantly exceeds our processor's $2^{68}$ Hilbert space dimension, and even the $\sim2^{265}$ atoms in the visible universe, which is surpassed near $N\sim10$.
\end{itemize}

\subsection{Optimizer parameters}
\label{sec:snakeparams}

\begin{figure*}[hbt!]
\includegraphics[width=\textwidth]{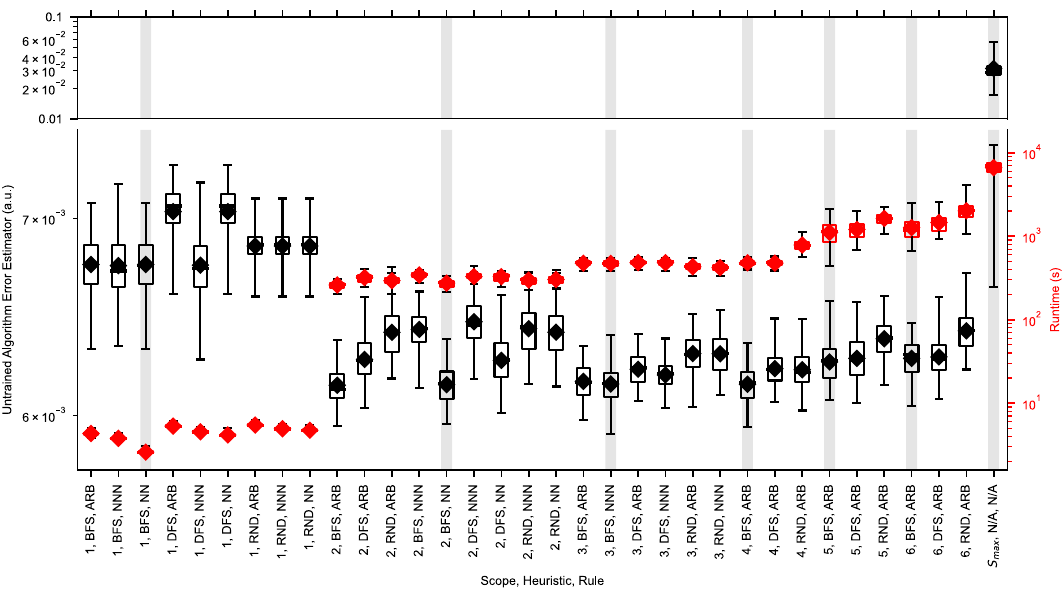}
\caption{\textbf{Tuning Snake's parameters}. The optimized untrained algorithm estimator value (black boxes) and runtime (red boxes) corresponding to $N=68$ optimization threads versus Snake's scope, traversal heuristic, and traversal rule parameters. Each box corresponds to multiple seeds and shows the respective distribution's 0, 25, 50, 75, and 100th percentile (horizontal notches) and mean (diamond). The gray vertical bars are the tuned parameters used in experiment for each scope.}
\label{fig:si_hyperparams}
\end{figure*}

\hlcyan{Snake can implement a wide array of optimization strategies through judicious selection of four parameters} - the seed strategy, traversal strategy, scope, and inner-loop optimizer. These must be tuned to the problem of interest, \hlcyan{while balancing runtime and performance, which often compete. We interpret this tuning as adapting Snake to the optimization landscape presented by the algorithm error estimator and expect it to be especially critical when applying our strategy beyond our hardware.}

Below we overview each parameter's role and then tune them over $\sim7,000$ simulated optimization threads of our quantum processor. We evaluate their runtime and performance against an untrained but representative algorithm error estimator, quoting relative performance measures (Figure \mbox{\ref{fig:si_hyperparams}}). By tuning the parameters against an estimator and not hardware benchmarks, these results are not susceptible to drift or the inaccuracy of the estimator in predicting benchmarks. Furthermore, an experiment of this scale would take $\sim 1.5$ years in hardware.

\hlcyan{
Once tuned, the parameters are remarkably robust. We have seen one set of tuned parameters remain reasonably effective for a variety of quantum algorithms and multiple generations of frequency tunable qubits, some of which underwent significant architectural modifications.
}

\subsubsection{Seed strategy}

Determines how Snake prioritizes seed gates to launch optimization threads from. 

\begin{itemize}
    \item \textbf{Performance impact}: $\sim 15\%$ variations. 
    \item \textbf{Runtime impact}: Negligible variations. 
    \item \textbf{Tuned value}: For experimental configurations up to $N=68$, we generated solutions from all seeds and selected the one that minimized the trained algorithm error estimator. For larger simulated processors, we selected a subset of all seeds randomly. 
\end{itemize}

We note that when scaling beyond hundreds of qubits, generating solutions from all seeds will become prohibitively time consuming. Identifying a strategy for seeding gates that may offer a performance advantage is thus expected to become important. One systematic approach for developing such a strategy is to correlate optimized estimator values against statistics computed from the corresponding seeds characterization data, which are available prior to optimization.  

\subsubsection{Traversal strategy}
Determines how Snake drives graph traversal within each optimization thread, which we characterize via a \textit{traversal rule} and \textit{traversal heuristic}: 

\begin{itemize}
\item \textbf{Traversal rule}: Identifies candidate gates for traversal, given the current gate. Here we considered nearest-neighbor (NN), next-nearest-neighbor (NNN), and arbitrary-scope (ARB) rules. ARB is a generalization of the NN and NNN rules that identifies candidates within an annular ring whose radius is set by the scope parameter. It was developed for $S>3$ optimization, where NN and NNN traversals are too short. 
\item \textbf{Traversal heuristic}: Sorts candidate gates returned by the traversal rule to implement some desired traversal pattern. Here we tested textbook breadth-first (BFS), depth-first (DFS), and random (RND) heuristics \cite{clrs}. 
\end{itemize}

\begin{itemize}
    \item \textbf{Performance impact}: $\sim 15\%$ variations. 
    \item \textbf{Runtime impact}: Negligible variations.
    \item \textbf{Tuned value}: We use a scope-dependent traversal strategy (gray bars in Figure \ref{fig:si_hyperparams}). However, the ARB traversal rule with the BFS traversal heuristic often outperformed other strategies. 
\end{itemize}

We note that many algorithms exist for approaching graph-based problems \cite{norvig, clrs} and can be adapted to Snake. For example, heuristics such as the fail-first heuristic (i.e. prioritize traversals between highly constrained gates) and/or techniques such as back-tracking \cite{norvig} and/or Monte Carlo Tree Search (MCTS) \cite{sutton, norvig} may offer advantages beyond the textbook strategies tested. \hlcyan{Multiple optimization traversals, which we consider an extension of healing, may also boost performance}.

\subsubsection{Scope}

\begin{figure*}[pht!]
\includegraphics[width=\textwidth]{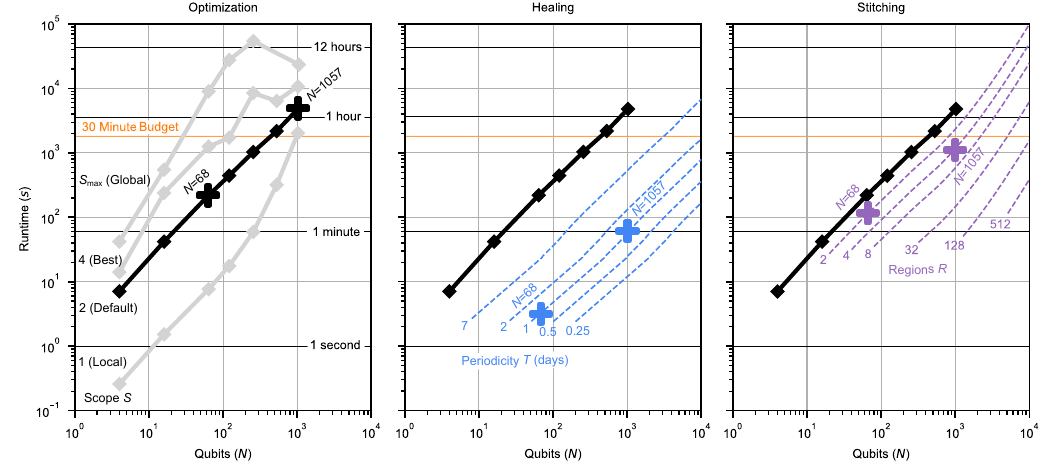}
\caption{\textbf{Optimization, healing, and stitching runtime scalability}. (a) Optimization runtime versus simulated processor size $N$ for several scopes $S$ of interest. The default scope $S=2$ (black) and 30 minute optimization budget (orange) are reproduced on all panels. \hlcyan{The ``+" markers are points of interest for the largest experimental ($N=68$) and simulated ($N=1057$) configurations investigated. (b) Estimated healing runtimes for several heal periodicities $T$. The ``+" markers correspond to daily healing. We extrapolate runtimes below the range of available data.} (c) Estimated stitching runtimes for several stitched regions $R$. \hlcyan{The ``+" markers correspond to the stitching experiment and simulation presented in the main text.}}
\label{fig:si_runtime_scalability}
\end{figure*}

Bounds the dimension of the optimization problem solved at each traversal step. We note that we refer to $S$ as the distance $d_P$ in reference \cite{snake} - we do not adopt that name here to avoid collisions with the error correction distance $d$. Here we tested scopes $S=1, 2, 3, 4, 5, 6, S_{\text{max}}$, which correspond to maximum optimization dimensions of 1D (local), 5D, 9D, 21D, 29D, 49D, $|F|$D (global). 

\begin{itemize}
    \item \textbf{Performance impact}: $\sim 5\times$ variations. 
    \item \textbf{Runtime impact}: $\sim10^4\times$ variations.
    \item \textbf{Tuned value}: $S=4$ ($\leq 21$D) found the lowest estimator values, when averaged over all tested hyperparameters. However, $S=2$ ($\leq 5$D) offers a better balance between performance (only $\sim 2 \%$ worse) and runtime ($\sim 50 \%$ faster) and is set as our tuned value. 
\end{itemize}

We note that the scope may be dynamically varied at each traversal step depending on the characterization data. For example, larger scopes may outperform lower scopes when optimizing highly constrained gates with anomalous circuit parameters and/or particularly detrimental TLS defects \cite{mullertls}. \hlcyan{Furthermore, we expect the optimal scope to depend strongly on the hardware architecture and to scale with the spatial extent of engineered and parasitic interactions. Optimizing hardware with higher connectivity (e.g. with three-qubit gates) would likely benefit from a larger scope than lower connectivity (e.g. with two-qubit gates). Similarly, optimizing hardware with longer-range stray coupling would benefit from a larger scope than shorter-range stray coupling.}

\subsubsection{Inner loop optimizer}
Determines how Snake optimizes the Snake estimator at each traversal step. Here we considered differential evolution \cite{de}, L-BFGS \cite{lbfgs}, simplical homology global optimization \cite{shgo}, dual annealing \cite{dualannealing}, dividing rectangles \cite{direct}, and basin hopping \cite{basinhopping}.  

\begin{itemize}
    \item \textbf{Performance impact}: Significant variations.
    \item \textbf{Runtime impact}: Significant variations.
    \item \textbf{Tuned value}: We use a dimension-dependent inner-loop optimizer. We exhaustively search $<3$D problems for their globally optimal values. We stochastically search $\geq 3$D problems via a tuned global optimizer.
\end{itemize}

We note that Snake can deploy nearly-arbitrary \hlcyan{continuous or discrete inner-loop optimizers, treating the optimization variables accordingly}. Of particular interest are model-free reinforcement-learning agents \cite{sutton, norvig, sivakrl, 21qctrlrl}. \hlcyan{In the short term, model-free agents could refine configurations found via model-based optimization to compensate for inaccuracies in the algorithm error estimator, which are expected to increase with processor size due to increased control and hardware inhomogeneities. In the longer term, they could replace model-based optimization entirely \mbox{\cite{20qctrl, hothem2023predictive}} and eliminate the research burden of developing performance estimators.}

\subsection{\hlcyan{Optimization runtime budget}}
\hlcyan{We develop a runtime budget with the objective of operating a distance 23 surface code logical qubit with $N=1057$ physical qubits. Since the surface code has a lenient qubit failure tolerance of $>1\%$ \mbox{\cite{qecdefect}} and since our outlier emergence probability is $\sim 0.01$ / 24 hours / gate (see Section \mbox{\ref{drift}}), we have $\sim24$ hours to characterize, optimize, calibrate, and benchmark our processor and finally execute the algorithm before restarting the process. 

If we target 12 hours for executing the algorithm, we have 12 hours to go from characterization to benchmarking. Since characterization, calibration, and benchmarking take $\sim2$ hours and are nominally independent of processor size due to parallelization, we have $\sim 10$ hours left. From this large window, we only budget 0.5 hours for optimization, leaving $\sim 9.5$ hours for unforeseen scaling overhead. 

The 0.5 hour runtime budget can be fulfilled via a combination of optimization, healing, and stitching (see Section \mbox{\ref{sec:runtime_scalability}}). Furthermore, as the hardware evolves, we expect the outlier emergence probability to decrease and the control system to become faster through hardware and software advancements, which should relax all budgets and enable operating even larger surface codes.}

\subsection{Optimization runtime scalability}
\label{sec:runtime_scalability}

To understand how optimization thread runtimes scale with processor size, we optimize simulated processors of variable size at several scopes (Figure \ref{fig:si_runtime_scalability}). The trends are complex, but runtimes generally increase with both scope and processor size. For our default scope $S=2$ in particular, runtimes fit well to the heuristic scaling model $r=a + bN + cN^2$, where $r$ is the runtime, $N$ is the number of qubits, and $a$, $b$, and $c$ are heuristic coefficients. 

Fitting this model to our data, we find best-fit parameters $a=-8.4 \pm 15.5$ seconds, $b=3.6 \pm 0.1$ seconds / qubit, and $c=0.0010 \pm 0.0001$ seconds / qubit$^2$. The linear term dominates towards hundreds of qubits, as we would expect from Snake's algorithmic structure. This is the term quoted in the scaling section of the main text. 

\hlcyan{To estimate healing runtimes $r_h$, we assume that outliers emerge with probability $0.01$ / 24 hours / gate (see Section \mbox{\ref{drift}}), that each outlier conservatively targets 2 qubits' idle and interaction frequencies for healing (see Section \mbox{\ref{healing}}), and that healing occurs periodically every $T$ days. In combination, $r_h = 0.01 \times 2 \times T \times r$.}

To estimate stitching runtimes $r_s$, we assume that a processor can be split into approximately equal-sized regions and that stitching overhead is negligible. We believe the latter assumption is valid, especially since stitching can itself be parallelized. In combination, $r_s = r / R$. 

\section{Benchmarking System}
\begin{figure}[t!]
\includegraphics[width=\linewidth]{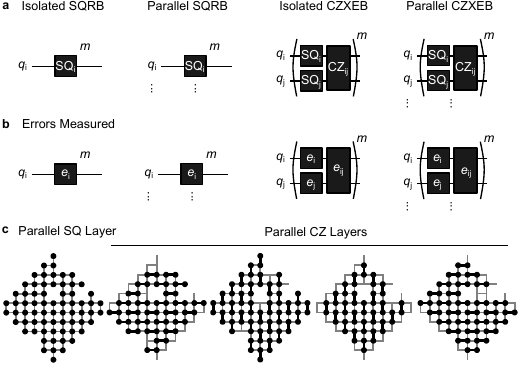}
\caption{\textbf{Benchmarking performance}. (a) Benchmark types used to evaluate performance. The ellipsis indicates that gates are benchmarked over more qubits in parallel. Initialization into $|0\rangle$ and measurement in the $Z$ basis are implicit. (b) Errors measured by the benchmark types in (a). (c) SQ and CZ gate \textit{layers} used for the ``Parallel" benchmarks in (a). Measuring CZXEB for all gates requires running four distinct algorithms, each of which is characterized by interleaving a distinct CZ layer with the SQ layer. When referencing CZXEB, we implicitly reference all four algorithms and corresponding benchmarks.}
\label{fig:si_benchmarks}
\end{figure}

\begin{figure}
\includegraphics[width=\linewidth]{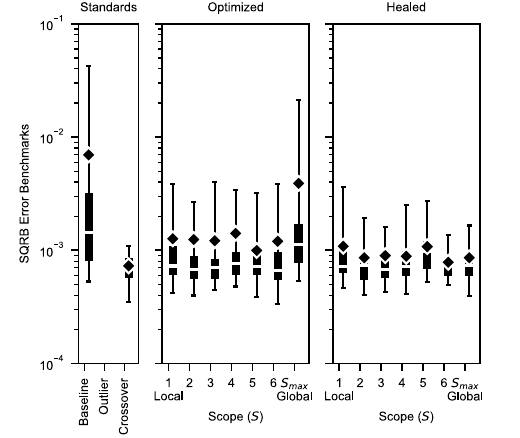}
\caption{\textbf{Optimization and healing performance}. SQRB benchmark distributions corresponding to the configurations presented in Figure 2. An outlier standard is not defined for SQRB. SQRB trends are consistent with CZXEB. }
\label{fig:si_sqrb_var_scope}
\end{figure}

\begin{figure}
\includegraphics[width=\linewidth]{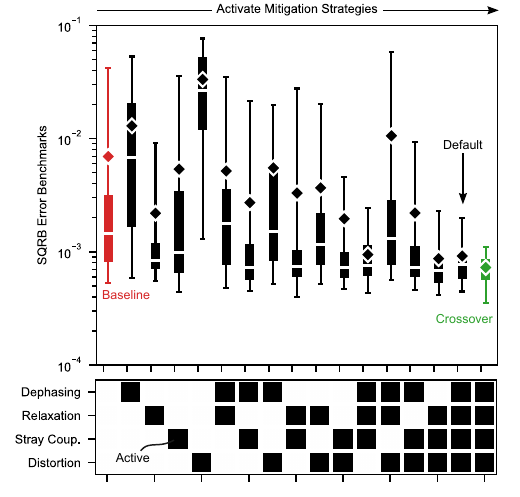}
\caption{\textbf{Optimization performance for various mitigation strategies}.
SQRB benchmarks distributions corresponding to the configurations presented in Figure 3.
An outlier standard is not defined for SQRB. SQRB trends are consistent with CZXEB.}
\label{fig:si_sqrb_var_mitig}
\end{figure}

\begin{figure}
\includegraphics[width=\linewidth]{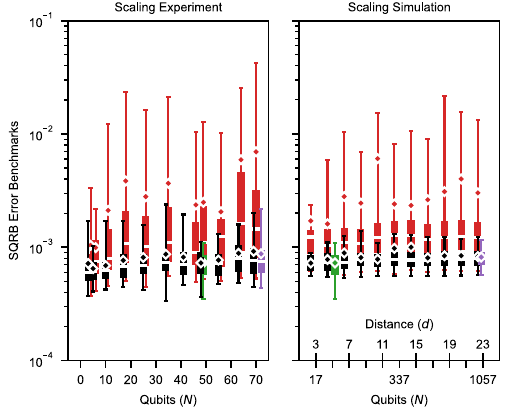}
\caption{\textbf{Optimization scalability}.
SQRB benchmark distributions corresponding to the configurations presented in Figure 4. An outlier standard is not defined for SQRB. SQRB trends are consistent with CZXEB. Some boxes have been horizontally shifted to reduce overlap.}
\label{fig:si_sqrb_scaling}
\end{figure}

The benchmarking system measures the performance of frequency configurations via benchmarking algorithms that should be representative of the target quantum algorithm but much cheaper to execute (Figure \ref{fig:si_benchmarks}). Here we use: 

\begin{itemize}

\item \textbf{Two-qubit cross-entropy benchmarking (CZXEB)}: Measures the error per cycle $e_{c,ij}$ for SQ$_i$, SQ$_j$ and CZ$_{ij}$, respectively \cite{boixo, sup1}. Four CZXEB benchmarking algorithms with distinct CZ ``Layer" patterns are necessary to benchmark all CZ gates (Figure \ref{fig:si_benchmarks}c). They are run back-to-back during benchmarking. When referencing ``CZXEB", we reference all of these algorithms and their respective benchmarks in combination.
\item \textbf{Single-qubit randomized benchmarking (SQRB)}: Measures the error per gate $e_i$ for SQ$_i$ (Figure \ref{fig:si_benchmarks}) \cite{Emerson_2005}. These benchmarks were not presented in the main text since CZXEB is considered a more holistic metric. SQRB benchmarks corresponding to the experiments in the main text are in Figures \ref{fig:si_sqrb_var_scope}, \ref{fig:si_sqrb_var_mitig}, and \ref{fig:si_sqrb_scaling}. SQRB can be combined with CZXEB to infer CZ error contributions to the cycle errors ($e_{ij}$ for CZ$_{ij}$).

\end{itemize}

In some cases, we label benchmarks as follows: 
\begin{itemize}
\item \textbf{Isolated}: Benchmarks taken in sparse configurations where stray coupling is negligible. 
\item \textbf{Parallel}: Benchmarks taken in denser configurations where stray coupling is non-negligible. Unless otherwise noted, all benchmarks are Parallel, including those presented in the main text. 
\end{itemize}

All SQRB and CZXEB benchmarks are reported as \textit{average errors}. Conversion factors between average errors, depolarizing errors, and Pauli errors are defined in Table I of Section V of the Supplementary information of Reference \cite{sup1}. Statistics for SQRB and CZXEB distributions presented in the main text and the this Supplementary information are in Tables \ref{tab:si_bench_var_scope}, \ref{tab:si_bench_var_mitig}, \ref{tab:si_bench_scaling_exp}, and \ref{tab:si_bench_scaling_sim}.   

\newpage

\vfill\null

\section{Simulation Environment}

\begin{figure*}[ht!]
\includegraphics[width=\textwidth]{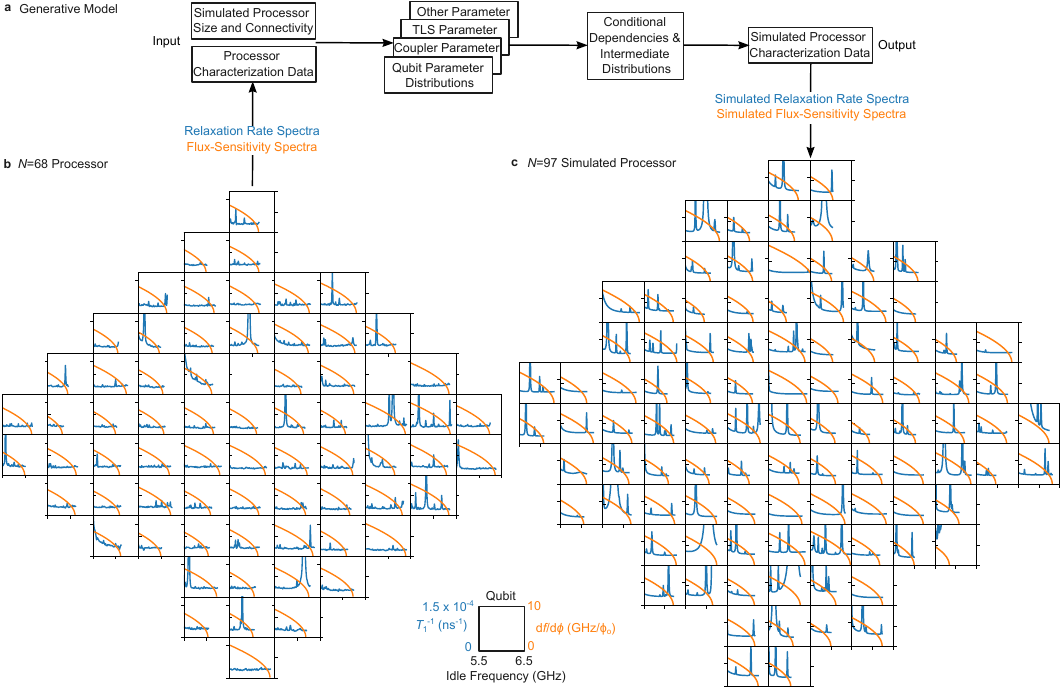}
\caption{\textbf{Generating simulated processors}. \hlcyan{(a) Schematic of the generative model that we use to generate simulated processor characterization data.} (b) Comparison of the energy relaxation rate ($T_1^{-1}$, blue) and flux-sensitivity spectra ($\frac{df}{d\phi}$, orange) for our $N=68$ processor and (b) $N=97$ simulated processor. The inset shows the common scale. Our generative model produces simulated processor characterization data that are nearly indistinguishable from our processor.}
\label{fig:si_fake_spectra}
\end{figure*}

Due to complex interplay between hardware inhomogeneities, error components, hard bounds, and our control optimization strategy, trustworthy scaling simulations require us to emulate our quantum computing stack from the hardware to the control system. We emulate them as follows: 

\begin{itemize}
\item \textbf{Hardware}: We developed \cite{gmpatent} a generative model of our quantum processor architecture (more below). This approach enables us to embed the statistics of empirically measured characterization data into our simulations.
\item \textbf{Characterization}: We sample \cite{norvig} the generative model to generate simulated processor characterization data. Each random sample is distinct but should nominally follow the statistics of empirically measured characterization data. 
\item \textbf{Optimization}: We use the Snake optimizer on simulated characterization data exactly as we would on real characterization data. 
\item \textbf{Calibration and Benchmarking}: We use our trained algorithm error estimator on optimized configurations with the simulated characterization data to estimate benchmarks. We assume that the impact of calibration is implicitly embedded into the trained weights of the estimator. 
\end{itemize}

Since the generative model is an important component of the simulation environment and has not yet been discussed in detail, we provide an overview below and direct the reader to Reference \cite{gmpatent} for additional details.

\subsection{Simulated processor generative model}
Our goal is to develop a statistical model that can be sampled for characterization data for simulated processors of arbitrary size and connectivity that are statistically indistinguishable from our real quantum processor. Towards that end, we interpret an arbitrary quantum processor as a statistical sample from some \textit{quantum processor joint probability density} $P(D, \mathcal{P}, \mathcal{N})$. Here $D$ is the characterization data, $\mathcal{P}$ is a set of architectural parameters (e.g. qubit circuit parameters, coupler circuit parameters, TLS parameters), and $\mathcal{N}$ is the processor's size and connectivity. Within this statistical picture, $D, \mathcal{P}, \mathcal{N}$ are random variables. 

Generating a simulated quantum processor and its corresponding characterization data amounts to sampling $P$. To do so, we consider the chain-rule decomposition $P(D, \mathcal{P}, \mathcal{N}) = P(D|\mathcal{P}, \mathcal{N})P(\mathcal{P}|\mathcal{N})P(\mathcal{N})$, which can be represented as a Bayesian network, and employ \textit{prior sampling} \cite{norvig} (Figure \ref{fig:si_fake_spectra}a) as follows: 

\begin{enumerate}
    \item Select simulated processor size and connectivity $\mathcal{N}$.
    \item Sample architectural parameters for all qubits of the simulated processor from $P(\mathcal{P}|\mathcal{N})$ under the naive assumption that they are conditionally independent amongst themselves and the qubits. The distributions of architectural parameters were determined by statistically analyzing our real quantum processor's characterization data.  
    \item Sample characterization data by propagating the architectural parameters through conditional dependencies $P(D|\mathcal{P}, \mathcal{N})$ determined through textbook physics, published literature, and metrology.
\end{enumerate}

If the generative model is accurate, the simulated processor's characterization data should be statistically indistinguishable from our real quantum processor. 

\subsection{Simulated characterization data}
We validate the accuracy of our generative model by comparing various statistics between the real and simulated architectural samples $\mathcal{P}$ and real and simulated characterization data $D$ (not shown). Here, we simply compare the relaxation and flux-sensitivity spectra sampled for a simulated processor against our real processor (Figure \ref{fig:si_fake_spectra}). We believe this is a holistic test of accuracy because these spectra have complex characteristics that require an accurate confluence of architectural parameters to accurately reproduce. Apart from experimental noise, which are filtered during optimization, the simulated spectra are nearly indistinguishable from real spectra. In turn, we believe the generative model is sufficiently accurate for trustworthy simulations.

\section{Additional Experimental Details}

\subsection{Experimental controls}
Due to the complexity of our quantum computing stack, developing good experimental controls for any sub-component - including our frequency optimization system - is non-trivial. Our primary controls are the random baseline configurations. These are expected to sample the average performance of the hardware and calibration systems without frequency optimization. However, we note that all baseline configurations are generated within frequency hard bounds, which themselves embed some error mitigation (see below). Therefore, we believe that the performance advantage quoted in the main text for our optimization strategy is an underestimate.  

\subsection{Impact of drift}
\label{drift}
Both slow gradual drift (e.g. due to temperature fluctuations in the lab) and abrupt catastrophic drift (e.g. due to TLS fluctuating into gates) can generate outliers over time. To understand the extent to which drift impacts our experimental results, we compare the runtimes of the components of our control system against estimated rate at which outliers emerge for $N=68$ configurations. 

\begin{itemize}
\item \textbf{Characterization}: $\sim 0.5$ hours / experiment. 
\item \textbf{Optimization}: $0.5$ hours budgeted / experiment. 
\item \textbf{Calibration}: $\sim 1$ hours / experiment. 
\item \textbf{Benchmarking}: $\sim ~0.2$ hours / experiment. 
\item \textbf{Total runtime}: $\sim 2.5$ hours / experiment.
\item \textbf{Outlier emergence probability}: \\$\sim 0.01$ / 24 hours / gate.
\end{itemize}

If we naively assume that outliers are conditionally independent and binomially distributed, we expect drift to generate:

\begin{align*}
\sim 177 \text{ gates} \times \frac{0.01}{24 \text{ hours} \times \text{gate}} \times \frac{2.5 \text{ hours}}{\text{experiment}} \sim \frac{0.2 \text{ outliers}}{\text{experiment}} 
\end{align*}

We thus do not believe that our results are significantly impacted by outliers emerging from drift. 

\subsection{Healing experiment}
\label{healing}
For the healing experiment presented in Figure 2, frequencies were targeted for healing via thresholding and manual discretion. 

\begin{itemize}
    \item Interaction frequencies were typically targeted when their corresponding CZXEB benchmarks exceeded the outlier standard ($e_{c,ij} \geq 1.5\times 10^{-2}$). Re-optimizing interactions is relatively inexpensive and does not necessarily require the corresponding idles to be re-optimized. 
    \item Idles were typically targeted when their corresponding SQRB benchmarks were anomalously high ($e_i\gtrsim 1.5\times 10^{-3}$) and/or when they corresponded to multiple CZXEB benchmarks that exceeded the outlier standard. Re-optimizing idles is relatively expensive and requires all hinged interactions to be re-optimized also.  
\end{itemize}

Heals were conducted within $\sim 1$ hour of benchmarking the optimized but unhealed configurations to mitigate the impact of drift. Furthermore, new characterization data were taken for targetted devices under the assumption that the original data were no longer valid. The primary concerns are TLS fluctuations \cite{klimtls, mullertls}, which are represented in the $T_1$ spectra.

\subsection{\hlcyan{Metrology experiment}}
\begin{figure*}[pth!]
\centering
\vspace*{2.5cm}
\includegraphics[width=1\textwidth]{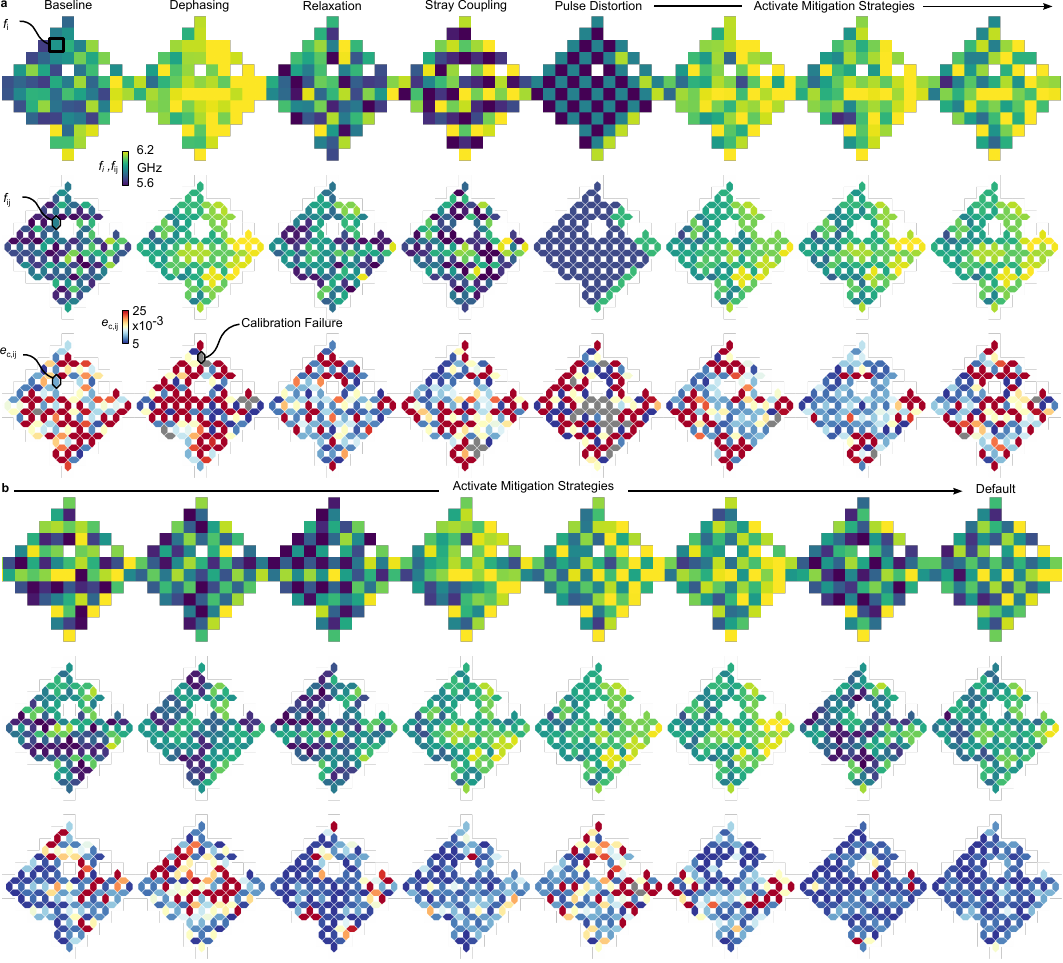}
\caption{\hlcyan{\textbf{Error metrology frequency configurations and benchmarks}. (a) (upper) Idle frequencies ($f_i$), (center) interaction frequencies ($f_{ij}$), and (lower) CZXEB cycle errors ($e_{c,ij}$) for configurations optimized with all combinations of error mitigation strategies activated, to supplement the data shown in Figure 3. (b) Continuation of (a), with all annotations shared.}}
\label{fig:si_all_heatmaps_vertical}
\vspace*{3cm}
\end{figure*}

\hlcyan{In Figure \mbox{\ref{fig:si_all_heatmaps_vertical}}, we show frequencies and cycle errors for all 16 configurations considered in the metrology experiment in Figure 3, each of which was optimized with distinct error mitigation strategies active. In the following sections, we qualitatively and quantitatively analyze these data to understand the impact of each error mitigation strategy and their interplay.}

\begin{figure*}[ht!]
\includegraphics[width=\linewidth]{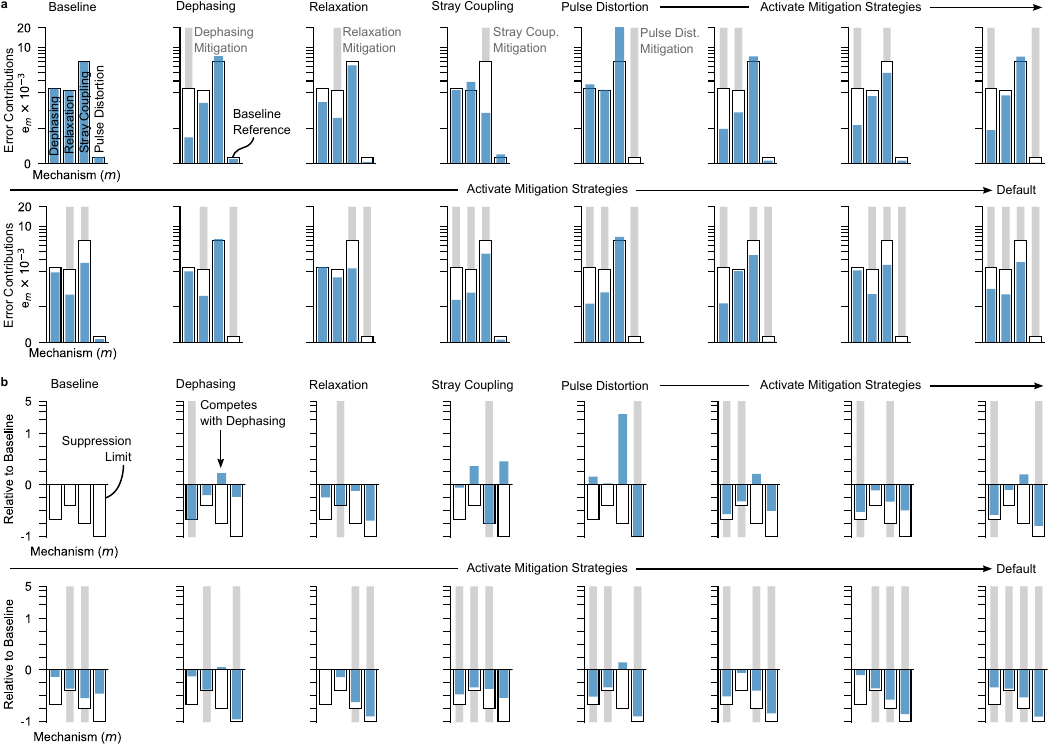}
\caption{\hlcyan{\textbf{Error metrology qualitative analysis}. (a) CZXEB cycle error contributions $e_m$ for each error mechanism $m$ and each configuration shown in Figure \mbox{\ref{fig:si_all_heatmaps_vertical}}. Each bar corresponds to one error mechanism and each panel corresponds to a distinct combination of error mitigation strategies active (grey). The baseline is reproduced on each panel for reference (white). (b) Error contributions from (a) relative to the baseline to highlight interactions between mitigation strategies. Amplification in an error mechanism when another mitigation strategy is active is interpreted as competition. For example, we show in the second panel that dephasing competes with stray coupling. Each error mechanism's suppression limit, which is achieved when only its respective mitigation strategy is active, is reproduced on each panel for reference (white). Ideally, our optimization strategy would reach each error mechanism's suppression limit when all mitigation strategies are active (last panel).}}
\label{fig:si_qualitative_errors}
\end{figure*}

\begin{figure*}[ht!]
\includegraphics[width=\linewidth]{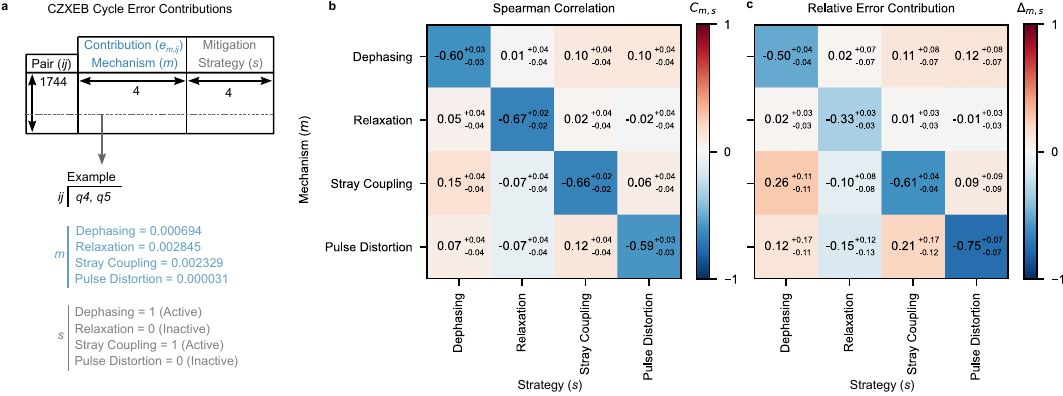}
\caption{\hlcyan{\textbf{Error metrology quantitative analysis}. (a) CZXEB cycle error contributions $e_{m,ij}$ for all error mechanisms $m$ and all pairs $ij$ across all 16 configurations, each of which corresponds to distinct error mitigation strategies $s$ active. One row is shown for example. (b) Spearman correlation $C_{m,s}$ between the activation of error mitigation strategy $s$ and error contributions from mechanism $m$. $C_{m,s} = -1$ means that activating strategy $s$ perfectly monotonically suppresses error mechanism $m$. Ideally, all diagonals would be $-1$ and all off-diagonals would be 0. (c) Relative error contribution $\Delta_{m,s}$ between the activation of error mitigation strategy $s$ and error contributions from mechanism $m$. $D_{m,s} = -1$ means that activating strategy $s$ fully suppresses mechanism $m$. Ideally, all diagonals would be $-1$ and all off-diagonals would be 0. For (b) and (c) we report medians and their 95\% confidence intervals.}}
\label{fig:si_error_mitigation_analysis}
\end{figure*}

\subsubsection{Qualitative analysis}
\hlcyan{
To qualitatively understand the impact of and interplay between error mitigation strategies, we inspect CZXEB cycle error contributions $e_m$ for all error mechanisms $m$ across all 16 configurations, each of which corresponds to distinct error mitigation strategies $s$ active (Figure \mbox{\ref{fig:si_qualitative_errors}}). Here $e_m = \langle e_{m, ij}\rangle_{ij}$, where $\langle.\rangle_{ij}$ is the median over pairs $ij$. The median was chosen to expose typical trends while preventing outliers from biasing our analysis. Each contribution $e_{m, ij}$ sums the corresponding weighted error components (e.g. SQ$_i$ dephasing, SQ$_j$ dephasing, and CZ$_{ij}$ dephasing for $m=$ dephasing) as computed from the corresponding configuration's optimized frequencies and characterization data. 

All panels in Figure \mbox{\ref{fig:si_qualitative_errors}} suggest that activating a particular error mitigation strategy suppresses its corresponding error mechanism as intended. However, the panels with only one mitigation strategy active also highlight non-trivial interactions between them. Of particular interest are amplifications in an error mechanism when another mitigation strategy is active, which we interpret as competition. Next we describe noteworthy competition and how it arises from the underlying physics.

Dephasing mitigation squeezes frequencies into a narrow band near their respective maxima where flux-sensitivity vanishes $\frac{df}{d\phi} = 0$ \mbox{\cite{07koch}} (see orange curves in Figure \mbox{\ref{fig:si_fake_spectra}}b). In turn, it boosts frequency collisions and thus competes with stray coupling. Stray coupling mitigation disperses frequencies to reduce frequency collisions. In turn, it boosts frequency excursions during CZs and thus competes with pulse distortion. Pulse distortion mitigation biases idles into $|11\rangle \leftrightarrow |02\rangle$ resonance (a checkerboard with neighbors $q_i$ and $q_j$ at $f_i = f_j - |\eta_j|$ or $f_j = f_i - |\eta_i|$, where $\eta_i$ and $\eta_j$ are qubit anharmonicities \mbox{\cite{dicarlo2009, surf}}) and interactions into resonance (at $f_{ij}=(f_i + f_j)/2$) to minimize frequency excursions during CZ gates.  In turn, it boosts frequency collisions and thus competes with stray coupling. Finally, relaxation mitigation interacts non-trivially with all mechanisms since it avoids relaxation hotspots with complex frequency dependencies \mbox{\cite{purcell_spontaneous_1946, 07koch, evanreadout, oliverboxmodes2021}} and randomness due to TLS \mbox{\cite{mullertls}} (see blue curves in Figure \mbox{\ref{fig:si_fake_spectra}}b). 

Despite nontrivial interactions between error mitigation strategies, the last panel of Figure \mbox{\ref{fig:si_qualitative_errors}}b suggests that our optimizer can effectively reconcile their competition and suppress all error mechanisms simultaneously. 
}
\subsubsection{Quantitative Analysis}
\hlcyan{
To quantitatively understand the impact of and interplay between error mitigation strategies, we analyze CZXEB cycle error contributions $e_{m,ij}$ for all error mechanisms $m$ and all pairs $ij$ across all 16 configurations, each of which corresponds to distinct error mitigation strategies $s$ active (Figure \mbox{\ref{fig:si_error_mitigation_analysis}}a). Towards that end, we construct a database of error contributions with 1744 rows (109 pairs $\times$ 16 configurations) and 16 columns (4 mechanisms $\times$ 4 strategies), randomly sample $\sim500$ rows, compute a metric of interest, and repeat 100,000 times. This bootstrapping procedure \mbox{\cite{elements, introstatlearning}} generates a distribution for the metric, from which we report the median and 95\% confidence interval. 

Our first metric is the Spearman correlation coefficient ($C_{m,s}$) between each error mitigation strategy and mechanism (Figure \mbox{\ref{fig:si_error_mitigation_analysis}}c). The Spearman correlation identifies monotonic trends, making it more suitable for analyzing non-linear data with outliers than the more common Pearson correlation. We find that activating a strategy is moderately-to-strongly correlated with a suppression in the corresponding error mechanism ($-0.67^{+0.02}_{-0.02} \leq C_{m,s} \leq -0.59^{+0.03}_{-0.03}$ for
diagonals) while being only weakly correlated with other mechanisms ($-0.07^{+0.04}_{-0.04} \leq C_{m,s} \leq 0.15^{+0.04}_{-0.04}$ for off-diagonals).

Our second metric is the relative error contribution $\Delta_{m, s} = e_{m,s=1}/e_{m,s=0} - 1$ (Figure \mbox{\ref{fig:si_error_mitigation_analysis}}d). Here $e_{m,s=1(0)} = \langle e_{m,ij}\rangle_{ij, s=1(0)}$ is the median over pairs in all configurations where strategy $s$ is active (inactive). We find that activating a particular strategy moderately-to-strongly suppresses its corresponding error mechanism ($-0.75^{+0.07}_{-0.07} \leq \Delta_{m,s} \leq -0.33^{+0.03}_{-0.03}$ for diagonals), while relatively weakly interacting with others ($-0.15^{+0.12}_{-0.13} \leq \Delta_{m,s} \leq 0.26^{+0.11}_{-0.11}$ for off-diagonals). Interestingly, relaxation mitigation appears to be less effective than other mitigation strategies ($\Delta_{m,s} = -0.33^{+0.03}_{-0.03}$ with $m=s=$relaxation). However, the last panel of Figure \mbox{\ref{fig:si_qualitative_errors}}b actually shows that relaxation approaches its suppression limit, even when all mitigation strategies are active, suggesting that relaxation is limited by our processor's performance limits and not our optimizer.

These results suggest that our mitigation strategies are both selective and effective at suppressing their corresponding error mechanisms. In turn, they support our association of error components with mitigation strategies and that our optimizer can effectively navigate them. 
}  

\subsection{Scaling experiment}
When building configurations for the scaling experiment presented in Figure 4, multiple configurations were used for smaller configuration sizes ($N<40$) to boost statistics (see ``Configurations" column in Table \ref{tab:si_bench_scaling_sim}). Optimized configurations were generally healed one or more times to resolve calibration failures and to push the performance limits of our processor. However, frequencies were never selected manually. Unoptimized baseline configurations were never healed. As a result, $\sim 1\%$ of all gates across all baseline configurations failed calibrations. 

The best-fit parameters for the saturation model are presented in the Table \ref{tab:si_sat_model}. The parameters $e_{sat}$ and $e_{scale}$ are remarkably similar between experiment and simulation. However, there is a sizeable gap for $N_{sat}$. We trust the experimental value more since the simulated data are sparse towards smaller $N$ (i.e. the $d=3, 5,$ and $7$ logical qubits have $N=7, 49,$ and $96$, respectively). 

\begin{table}
\centering
\caption{Best-fit parameters for the saturation model introduced in the main text. The error bars are standard deviations ($\pm 1\sigma$).}
\begin{tabular}{c|c|c|c|c}
\hline
 &\multicolumn{2}{c|}{Experiment} & \multicolumn{2}{c}{Simulation} \\
\hline
  & Baseline & Optimized & Baseline & Optimized \\
\hline
$N_{sat}$ & $21 \pm 21$ &  $22 \pm 10$ &  $49 \pm 46$ &  $84 \pm 41$ \\
$e_{sat} \times 10^{-3}$ & $28 \pm 5$ & $7.5 \pm 0.4$ & $24 \pm 1$ & $7.5 \pm 0.2$ \\
$e_{scale} \times 10^{-3}$ & $17 \pm 5$ & $3.1 \pm 0.4$ & $12 \pm 6$ & $2.2 \pm 0.6$ \\
\end{tabular}
\label{tab:si_sat_model}
\end{table}

\subsection{\hlcyan{Stitching experiment}}
\hlcyan{The stitching demonstrations in Figure 4 employed convenient stitch geometries. Even though stitched-and-healed configurations performed as well as their unstitched counterparts, we expect that the number of stitched regions and seam geometry will ultimately need to be optimized. First, we expect that progressively increasing the number of stitched regions - which would favorably lead to shorter optimization runtimes - will eventually start to degrade performance as more constraints between more independently optimized configurations will have to be reconciled. Second, we expect that optimizing the seam geometry will be necessary for applications like error correction, where the geometry of underperformant gates is particularly important \mbox{\cite{fowler2012, qecdefect}}}.

\subsection{Algorithm specificity}

\begin{figure}[ht!]
\includegraphics[width=\linewidth]{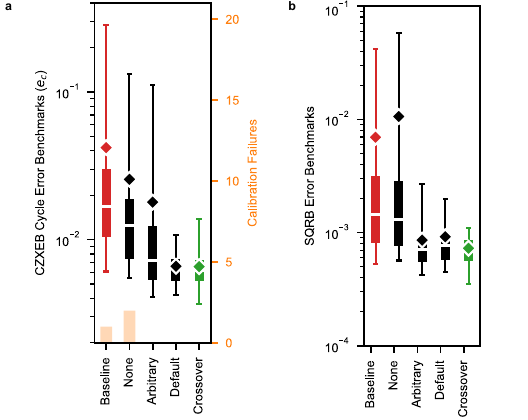}
\caption{\textbf{Impact of algorithm-specific optimization}.
(a) CZXEB cycle error and \hlcyan{(b) SQRB error benchmarks} for configurations optimized without stray coupling mitigation (``None"), with arbitrary-algorithm stray coupling mitigation (``Arbitrary"), and the default CZXEB-specific stray coupling mitigation (``Default"). The baseline (red), outlier (orange), and crossover standards are shown for comparison. Calibration failures are red bars (right axis).}
\label{fig:si_alg_specific}
\end{figure}

Ideally a single frequency configuration could reach high performance on arbitrary quantum algorithms. However, that configuration would have to mitigate frequency collisions between all possible combinations of parasitically coupled gates. That stringent requirement may be infeasible from a constraint perspective. 

To test this possibility, we optimize our processor for arbitrary quantum algorithms by configuring the stray coupling error components to penalize for frequency collisions between all possible combinations of parasitically coupled SQ and CZ gates. We compare the performance of the configuration against configurations optimized without stray coupling mitigation, with the default CZXEB-specific stray coupling mitigation, and the baseline, outlier, and crossover standards.

The configuration optimized for arbitrary quantum algorithms significantly underperforms the default configuration optimized with CZXEB-specific stray coupling mitigation and the crossover standard. Furthermore, it only moderately outperforms the configuration without stray coupling mitigation at all and the random baseline standard (Figure \ref{fig:si_alg_specific}). This result suggests that the optimization problem is overconstrained by the added stray coupling error components. Therefore, with the current magnitude of stray coupling, algorithm-specific optimization is critical.  

\section{Dependencies}
Optimization was performed on an Lenovo ThinkStation P620 with 128GB of ram and an AMD Ryzen Threadripper PRO 3945WX. Data were analyzed using a combination of \texttt{numpy} \cite{numpy}, \texttt{scipy} \cite{scipy}, \texttt{pandas} \cite{pandas}, and \texttt{matplotlib} \cite{matplotlib}. Boxplots were generated using \texttt{matplotlib.boxplot}. The percentiles in Tables \ref{tab:si_bench_var_scope}, \ref{tab:si_bench_var_mitig}, \ref{tab:si_bench_scaling_exp}, \ref{tab:si_bench_scaling_sim} were generated using \texttt{numpy.percentile}. The tested inner-loop optimizers leveraged a mixture of homebrew code and \texttt{scipy}'s built-in optimizers. The training pipeline was written in \texttt{tensorflow} \cite{tensorflow}.

\begin{table*}
\centering
\caption{Benchmarks ($\times 10^{-3}$) corresponding to the distributions shown in Figure 2 of the main text.}
\begin{tabular}{l|l|l|c|rrr|rrrrr}
\hline
  Benchmark & $N$ & Label & Scope ($S$) & Min &  Max &  Mean &  2.5\% &  25\% &  50\% &  75\% &  97.5\% \\
\hline
CZXEB & 68 & Baseline & - &    3.7 &      300.0 &        42.0 &          6.1 &        10.7 &        16.7 &        29.4 &         283.8 \\
CZXEB & - & Outlier & - & - &    - &     15.0 &      - &     - &     - &     - &       - \\ 
CZXEB & 49 & Crossover & - &    2.5 &       14.8 &         6.6 &          3.7 &         5.3 &         6.2 &         7.2 &          13.8 \\
\hline
CZXEB & 68 & Optimized & 1 &        2.5 &      300.0 &        25.2 &          3.7 &         7.4 &         9.8 &        15.5 &         241.6 \\
CZXEB & 68 & Optimized & 2 &        4.0 &       64.3 &         9.2 &          4.7 &         5.7 &         7.2 &         9.3 &          27.0 \\
CZXEB & 68 & Optimized & 3 &        3.7 &       29.4 &         9.0 &          4.8 &         6.3 &         7.4 &         9.8 &          20.9 \\
CZXEB & 68 & Optimized & 4 &        3.9 &       31.4 &         8.5 &          4.1 &         5.5 &         6.8 &         9.5 &          26.1 \\
CZXEB & 68 & Optimized & 5 &        2.5 &      295.0 &        20.4 &          3.9 &         6.3 &         7.6 &        12.2 &         136.4 \\
CZXEB & 68 & Optimized & 6 &        2.5 &      181.5 &        20.2 &          4.2 &         6.4 &         8.4 &        14.8 &         122.7 \\
CZXEB & 68 & Optimized & $S_{\text{max}}$ &        3.5 &      299.9 &        27.1 &          5.2 &         8.6 &        10.8 &        19.2 &         155.8 \\
\hline
CZXEB & 68 & Healed & 1 &        4.6 &       49.6 &        10.6 &          5.0 &         7.5 &         8.7 &        11.6 &          23.5 \\
CZXEB & 68 & Healed & 2 &        3.9 &       54.6 &         8.6 &          4.4 &         6.0 &         7.2 &         9.0 &          20.5 \\
CZXEB & 68 & Healed & 3 &        3.7 &       91.2 &         9.2 &          4.4 &         6.2 &         7.7 &         9.1 &          22.6 \\
CZXEB & 68 & Healed & 4 &        2.7 &       62.8 &         8.5 &          4.0 &         5.6 &         6.9 &         8.6 &          18.8 \\
CZXEB & 68 & Healed & 5 &        2.5 &       53.6 &         8.7 &          4.3 &         5.6 &         7.3 &         9.3 &          18.9 \\
CZXEB & 68 & Healed & 6 &        2.5 &      275.2 &        12.7 &          3.8 &         5.6 &         6.8 &         9.9 &          31.3 \\
CZXEB & 68 & Healed & $S_{\text{max}}$&        3.5 &      124.4 &        14.0 &          4.5 &         6.8 &         8.8 &        13.1 &          66.7 \\
\hline
\hline
SQRB & 49 &     Baseline & - &   0.43 &     53.19 &       6.96 &        0.53 &       0.83 &       1.45 &       3.07 &        42.02 \\
SQRB & - &     Outlier & - &  - &    - &     - &      - &     - &     - &     - &       - \\ 
SQRB & 49 &     Crossover & - &   0.33 &      1.47 &       0.73 &        0.35 &       0.58 &       0.72 &       0.84 &         1.10 \\
\hline
SQRB & 68 & Optimized & 1 &      0.40 &     12.75 &       1.26 &        0.42 &       0.61 &       0.73 &       1.04 &         3.86 \\
SQRB & 68 & Optimized & 2 &      0.26 &     25.74 &       1.24 &        0.40 &       0.58 &       0.68 &       0.87 &         2.67 \\
SQRB & 68 & Optimized & 3 &      0.36 &     13.87 &       1.21 &        0.44 &       0.58 &       0.70 &       0.82 &         3.99 \\
SQRB & 68 & Optimized & 4 &      0.39 &     27.49 &       1.41 &        0.48 &       0.62 &       0.76 &       0.94 &         3.41 \\
SQRB & 68 & Optimized & 5 &      0.36 &      4.61 &       1.00 &        0.38 &       0.61 &       0.72 &       1.03 &         3.22 \\
SQRB & 68 & Optimized & 6 &      0.00 &     11.14 &       1.20 &        0.34 &       0.56 &       0.67 &       0.94 &         3.82 \\
SQRB & 68 & Optimized & $S_{\text{max}}$ &      0.38 &     91.07 &       3.90 &        0.53 &       0.80 &       1.12 &       1.65 &        21.30 \\
\hline
SQRB & 68 & Healed & 1 &      0.38 &      7.80 &       1.09 &        0.47 &       0.64 &       0.72 &       1.06 &         3.62 \\
SQRB & 68 & Healed & 2 &      0.34 &      4.11 &       0.86 &        0.40 &       0.56 &       0.74 &       0.86 &         1.94 \\
SQRB & 68 & Healed & 3 &      0.40 &      9.92 &       0.90 &        0.43 &       0.58 &       0.68 &       0.80 &         1.60 \\
SQRB & 68 & Healed & 4 &      0.32 &      2.82 &       0.88 &        0.41 &       0.61 &       0.72 &       0.96 &         2.51 \\
SQRB & 68 & Healed & 5 &      0.49 &      4.10 &       1.08 &        0.53 &       0.70 &       0.97 &       1.18 &         2.73 \\
SQRB & 68 & Healed & 6 &      0.04 &      1.93 &       0.78 &        0.49 &       0.61 &       0.75 &       0.85 &         1.36 \\
SQRB & 68 & Healed & $S_{\text{max}}$ &      0.29 &      4.72 &       0.86 &        0.40 &       0.61 &       0.73 &       0.90 &         1.66 \\
\hline
\end{tabular}
\label{tab:si_bench_var_scope}
\end{table*}

\begin{table*}
\centering
\caption{Benchmarks ($\times 10^{-3}$) corresponding to the distributions shown in Figure 3 of the main text.}
\begin{tabular}{l|l|llll|rrr|rrrrr}
\hline
    Benchmark & $N$ & Dephasing &  Relaxation &  Stray coupling &  Pulse distortion & Min &  Max &  Mean &  2.5\% &  25\% &  50\% &  75\% &  97.5\% \\
\hline
    CZXEB & - &\multicolumn{4}{c|}{Outlier} &        - &    - &     15.0 &      - &     - &     - &     - &       - \\ 
    CZXEB & 68 &\multicolumn{4}{c|}{Baseline} &        3.7 &      300.0 &        42.0 &          6.1 &        10.7 &        16.7 &        29.4 &         283.8 \\
\hline
    CZXEB & 68 &1 &    0 &    0 &    0 &        2.5 &      299.9 &        42.2 &          2.7 &         7.5 &        21.5 &        43.1 &         222.2 \\
    CZXEB & 68 &0 & 1 &    0 &    0 &        2.5 &      285.1 &        20.7 &          3.7 &         7.4 &         9.3 &        18.2 &          92.7 \\
    CZXEB & 68 &0 &    0 & 1 &    0 &        2.5 &      283.2 &        29.9 &          3.9 &         8.2 &        14.8 &        29.4 &         192.1 \\
    CZXEB & 68 &0 &    0 &    0 & 1 &        2.5 &      299.8 &        53.2 &          2.5 &         3.8 &        15.4 &        54.7 &         266.2 \\
    CZXEB & 68 &1 & 1 &    0 &    0 &        2.5 &      277.6 &        32.1 &          5.2 &         8.2 &        13.4 &        29.4 &         184.1 \\
    CZXEB & 68 &1 &    0 & 1 &    0 &        3.7 &      287.4 &        19.1 &          4.1 &         7.2 &         9.2 &        12.8 &          73.8 \\
    CZXEB & 68 &1 &    0 &    0 & 1 &        3.7 &      237.3 &        28.5 &          3.7 &         8.0 &        14.8 &        30.1 &         127.7 \\
    CZXEB & 68 &0 & 1 & 1 &    0 &        2.5 &      286.7 &        14.9 &          4.2 &         5.9 &         8.0 &        14.2 &          46.9 \\
    CZXEB & 68 &0 & 1 &    0 & 1 &        2.5 &      299.7 &        25.7 &          3.7 &         7.4 &        12.6 &        18.8 &         124.9 \\
    CZXEB & 68 &0 &    0 & 1 & 1 &        2.5 &      139.9 &        12.1 &          4.0 &         5.2 &         6.3 &         9.2 &          52.7 \\
    CZXEB & 68 &1 & 1 & 1 &    0 &        3.7 &       27.3 &         8.1 &          4.2 &         5.8 &         7.4 &         9.3 &          16.5 \\
    CZXEB & 68 &1 & 1 &    0 & 1 &        3.7 &      299.0 &        25.6 &          5.5 &         7.5 &        12.4 &        18.0 &         132.6 \\
    CZXEB & 68 &1 &    0 & 1 & 1 &        3.6 &      224.0 &        15.1 &          3.7 &         6.0 &         8.2 &        12.4 &          79.0 \\
    CZXEB & 68 &0 & 1 & 1 & 1 &        3.8 &      102.4 &         8.2 &          4.2 &         5.2 &         6.2 &         7.3 &          24.8 \\
    CZXEB & 68 &1 & 1 & 1 & 1 &        4.0 &       14.5 &         6.6 &          4.2 &         5.3 &         6.4 &         7.3 &          10.7 \\
\hline
    CZXEB & 49 &\multicolumn{4}{c|}{Crossover} &        2.5 &       14.8 &         6.6 &          3.7 &         5.3 &         6.2 &         7.2 &          13.8 \\
\hline
\hline
    SQRB & - &\multicolumn{4}{c|}{Outlier} &        - &    - &     - &      - &     - &     - &     - &       - \\ 
    SQRB & 68 &\multicolumn{4}{c|}{Baseline} &      0.43 &     53.19 &       6.96 &        0.53 &       0.83 &       1.45 &       3.07 &        42.02 \\
\hline
    SQRB & 68 &1 &    0 &    0 &    0 &      0.05 &     74.55 &      12.99 &        0.59 &       1.70 &       6.86 &      20.22 &        53.15 \\
    SQRB & 68 &0 & 1 &    0 &    0 &      0.53 &     38.31 &       2.19 &        0.55 &       0.72 &       0.84 &       1.17 &         9.16 \\
    SQRB & 68 &0 &    0 & 1 &    0 &      0.01 &     46.63 &       5.37 &        0.44 &       0.67 &       0.99 &       3.02 &        35.67 \\
    SQRB & 68 &0 &    0 &    0 & 1 &      0.08 &    188.21 &      33.32 &        1.30 &      12.50 &      26.52 &      50.78 &        76.57 \\
    SQRB & 68 &1 & 1 &    0 &    0 &      0.45 &     44.92 &       5.17 &        0.48 &       0.79 &       1.79 &       3.39 &        34.88 \\
    SQRB & 68 &1 &    0 & 1 &    0 &      0.40 &     61.34 &       2.72 &        0.45 &       0.59 &       0.73 &       1.12 &        21.52 \\
    SQRB & 68 &1 &    0 &    0 & 1 &      0.47 &     65.69 &       5.51 &        0.52 &       0.86 &       1.50 &       5.33 &        19.94 \\
    SQRB & 68 &0 & 1 & 1 &    0 &      0.02 &     38.28 &       3.30 &        0.40 &       0.61 &       0.74 &       1.01 &        27.70 \\
    SQRB & 68 &0 & 1 &    0 & 1 &      0.47 &     51.42 &       3.67 &        0.52 &       0.78 &       1.17 &       2.15 &        20.09 \\
    SQRB & 68 &0 &    0 & 1 & 1 &      0.29 &     61.05 &       1.96 &        0.47 &       0.61 &       0.73 &       0.89 &         4.59 \\
    SQRB & 68 &1 & 1 & 1 &    0 &      0.41 &      3.95 &       0.95 &        0.43 &       0.63 &       0.76 &       1.12 &         2.43 \\
    SQRB & 68 &1 & 1 &    0 & 1 &      0.47 &    199.96 &      10.60 &        0.57 &       0.78 &       1.30 &       2.79 &        58.24 \\
    SQRB & 68 & 1 &    0 & 1 & 1 &      0.01 &     33.36 &       2.20 &        0.46 &       0.61 &       0.73 &       1.08 &         9.42 \\
    SQRB & 68 &0 & 1 & 1 & 1 &      0.29 &      7.45 &       0.87 &        0.42 &       0.54 &       0.69 &       0.82 &         2.28 \\
    SQRB & 68 &1 & 1 & 1 & 1 &      0.34 &      4.64 &       0.92 &        0.45 &       0.59 &       0.77 &       0.96 &         1.99 \\
\hline
    SQRB & 49 &\multicolumn{4}{c|}{Crossover} &      0.33 &      1.47 &       0.73 &        0.35 &       0.58 &       0.72 &       0.84 &         1.10 \\
\hline
\end{tabular}
\label{tab:si_bench_var_mitig}
\end{table*}

\begin{table*}
\centering
\caption{Benchmarks ($\times 10^{-3}$) corresponding to the distributions shown in Figure 4 of the main text.}
\begin{tabular}{l|l|c|l|rrr|rrrrr}
\hline
Benchmark & $N$ & Configurations & Label & Min &  Max &  Mean &  2.5\% &  25\% &  50\% &  75\% &  97.5\% \\
\hline
CZXEB & - & - & Outlier &   - &    - &     15.0 &      - &     - &     - &     - &       - \\ 
\hline
CZXEB & 2  & 94& Optimized &         2.7 &       23.6 &         5.0 &          3.0 &         3.7 &         4.3 &         5.4 &          10.2 \\
CZXEB & 4  & 39& Optimized &        2.8 &       17.7 &         4.9 &          2.9 &         3.8 &         4.5 &         5.3 &           9.0 \\
CZXEB & 9  & 15& Optimized &        2.5 &       11.6 &         5.0 &          3.4 &         3.9 &         4.5 &         5.6 &           8.5 \\
CZXEB & 16 & 5& Optimized &        3.2 &       15.6 &         5.9 &          3.5 &         4.7 &         5.4 &         6.8 &          10.2 \\
CZXEB & 24 & 6& Optimized &        3.1 &       31.9 &         6.5 &          3.6 &         4.5 &         5.8 &         7.2 &          14.9 \\
CZXEB & 33 & 2& Optimized &        3.2 &       16.7 &         6.7 &          3.9 &         4.8 &         6.0 &         7.5 &          14.4 \\
CZXEB & 40 & 1& Optimized &        3.9 &       16.5 &         7.4 &          4.3 &         5.6 &         6.8 &         8.9 &          12.9 \\
CZXEB & 47 & 1& Optimized &        3.7 &       15.2 &         7.5 &          4.3 &         5.6 &         6.8 &         8.7 &          14.8 \\
\hline
CZXEB & 49 & -& Crossover & 2.5 &       14.8 &         6.6 &          3.7 &         5.3 &         6.2 &         7.2 &          13.8 \\
\hline
CZXEB & 54 & 1& Optimized &        3.7 &       29.4 &         7.4 &          4.4 &         5.8 &         6.9 &         8.1 &          13.7 \\
CZXEB & 62 & 1& Optimized &        3.7 &       17.9 &         7.6 &          4.7 &         6.0 &         7.2 &         8.4 &          12.8 \\
CZXEB & 68 & 1& Optimized &        4.0 &       14.5 &         6.6 &          4.2 &         5.3 &         6.4 &         7.3 &          10.7 \\
\hline
CZXEB & 68 & 1& Stitched ($R=2$) &        4.2 &       20.6 &         7.0 &          4.6 &         5.5 &         6.4 &         7.7 &          10.8 \\
\hline
CZXEB & 2  & 109& Baseline &       2.2 &      157.1 &        15.5 &          3.4 &         4.6 &         5.8 &        10.2 &          95.5 \\
CZXEB & 4  & 37& Baseline &         3.4 &       97.6 &         9.5 &          3.9 &         4.9 &         6.8 &         9.1 &          34.4 \\
CZXEB & 9  & 18& Baseline &         3.0 &      300.0 &        15.2 &          4.2 &         6.2 &         7.8 &        12.7 &          85.7 \\
CZXEB & 16 & 6& Baseline &         2.5 &      291.4 &        25.1 &          4.7 &         7.5 &        10.8 &        18.1 &         135.9 \\
CZXEB & 24 & 7& Baseline &         3.1 &      299.1 &        20.2 &          5.1 &         8.1 &        12.5 &        18.2 &          82.1 \\
CZXEB & 33 & 3& Baseline &         2.5 &      295.7 &        26.9 &          3.7 &         7.9 &        14.0 &        28.2 &         124.0 \\
CZXEB & 44 & 1& Baseline &         2.5 &      184.0 &        24.2 &          3.7 &         8.3 &        13.9 &        25.4 &         126.6 \\
CZXEB & 47 & 1& Baseline &         4.0 &      293.4 &        24.4 &          5.3 &         9.4 &        14.8 &        26.7 &          75.6 \\
CZXEB & 54 & 1& Baseline &         2.5 &      288.9 &        21.0 &          3.7 &         7.6 &        11.6 &        14.8 &         102.4 \\
CZXEB & 62 & 1& Baseline &         2.5 &      277.8 &        33.4 &          6.7 &        13.0 &        16.8 &        29.4 &         221.4 \\
CZXEB & 68 & 1& Baseline &         3.7 &      300.0 &        42.0 &          6.1 &        10.7 &        16.7 &        29.4 &         283.8 \\
\hline
SQRB & - & -& Outlier &        - &    - &     - &      - &     - &     - &     - &       - \\ 
\hline
SQRB & 2  & 94& Optimized &      0.29 &      6.43 &       0.72 &        0.37 &       0.51 &       0.60 &       0.71 &         1.72 \\
SQRB & 4  & 39& Optimized &      0.35 &      3.07 &       0.65 &        0.41 &       0.52 &       0.61 &       0.72 &         1.03 \\
SQRB & 9  & 15& Optimized &      0.31 &      2.48 &       0.69 &        0.42 &       0.54 &       0.62 &       0.72 &         1.72 \\
SQRB & 16 & 5& Optimized &      0.36 &      2.02 &       0.77 &        0.44 &       0.55 &       0.67 &       0.85 &         1.69 \\
SQRB & 24 & 6& Optimized &      0.32 &      7.14 &       0.81 &        0.42 &       0.59 &       0.69 &       0.86 &         1.58 \\
SQRB & 33 & 2& Optimized &      0.24 &      2.84 &       0.87 &        0.34 &       0.63 &       0.73 &       0.84 &         2.38 \\
SQRB & 40 & 1& Optimized &      0.41 &      2.10 &       0.82 &        0.46 &       0.58 &       0.70 &       0.89 &         1.94 \\
SQRB & 47 & 1& Optimized &      0.22 &      1.44 &       0.73 &        0.36 &       0.60 &       0.71 &       0.88 &         1.12 \\
\hline
SQRB & 49 & -& Crossover &      0.33 &      1.47 &       0.73 &        0.35 &       0.58 &       0.72 &       0.84 &         1.10 \\
\hline
SQRB & 54 & 1& Optimized &      0.31 &      1.62 &       0.77 &        0.48 &       0.58 &       0.71 &       0.85 &         1.30 \\
SQRB & 62 & 1& Optimized &      0.37 &      2.74 &       0.88 &        0.48 &       0.64 &       0.80 &       1.00 &         1.59 \\
SQRB & 68 & 1& Optimized &      0.34 &      4.64 &       0.92 &        0.45 &       0.59 &       0.77 &       0.96 &         1.99 \\
\hline
SQRB & 68 & 1& Stitched ($R=2$) &      0.41 &      3.23 &       0.88 &        0.44 &       0.61 &       0.75 &       0.93 &         2.16 \\
\hline
SQRB & 2  & 109& Baseline &      0.28 &     21.74 &       1.04 &        0.37 &       0.56 &       0.65 &       0.85 &         3.36 \\
SQRB & 4  & 37& Baseline &      0.28 &      5.87 &       0.99 &        0.41 &       0.59 &       0.75 &       1.14 &         2.19 \\
SQRB & 9  & 18& Baseline &      0.32 &     50.40 &       2.12 &        0.46 &       0.65 &       0.79 &       1.41 &        12.15 \\
SQRB & 16 & 6& Baseline &      0.37 &     60.85 &       3.85 &        0.50 &       0.75 &       1.08 &       2.06 &        23.35 \\
SQRB & 24 & 7& Baseline &      0.30 &     53.72 &       2.80 &        0.45 &       0.70 &       1.02 &       1.85 &        16.22 \\
SQRB & 33 & 3& Baseline &      0.43 &     46.46 &       3.66 &        0.56 &       0.77 &       1.10 &       2.14 &        21.29 \\
SQRB & 44 & 1& Baseline &      0.01 &     16.37 &       2.37 &        0.49 &       0.72 &       0.90 &       1.55 &        10.36 \\
SQRB & 47 & 1& Baseline &      0.00 &     18.90 &       2.49 &        0.52 &       0.81 &       1.10 &       2.17 &        12.68 \\
SQRB & 54 & 1& Baseline &      0.30 &     25.93 &       2.06 &        0.50 &       0.72 &       1.24 &       1.74 &        10.16 \\
SQRB & 62 & 1& Baseline &      0.48 &     76.72 &       5.93 &        0.53 &       0.96 &       1.61 &       4.40 &        25.50 \\
SQRB & 68 & 1& Baseline &      0.43 &     53.19 &       6.96 &        0.53 &       0.83 &       1.45 &       3.07 &        42.02 \\
\hline
\end{tabular}
\label{tab:si_bench_scaling_exp}
\end{table*}

\begin{table*}
\centering
\caption{Benchmarks ($\times 10^{-3}$) corresponding to the distributions shown in Figure 4 of the main text.}
\begin{tabular}{l|l|l|c|l|rrr|rrrrr}
\hline
Benchmark & $N$ & Distance & Configurations &     Label & Min &  Max &  Mean &  2.5\% &  25\% &  50\% &  75\% &  97.5\% \\
\hline
CZXEB & - & - & - & Outlier &   - &    - &     15.0 &      - &     - &     - &     - &       - \\ 
\hline
CZXEB & 17   &        3 &   1&Optimized &        3.3 &       10.0 &         5.5 &          3.8 &         4.6 &         5.6 &         6.0 &           7.2 \\
CZXEB & 49   &        5 &   1&Optimized &        3.4 &       12.6 &         6.3 &          3.6 &         5.1 &         6.1 &         7.3 &          10.6 \\
\hline
CZXEB & 49   &        5 &   1&Crossover & 2.5 &       14.8 &         6.6 &          3.7 &         5.3 &         6.2 &         7.2 &          13.8 \\
\hline
CZXEB & 97   &        7 &   1&Optimized &        3.1 &       25.3 &         7.2 &          3.8 &         5.4 &         6.9 &         8.3 &          11.9 \\
CZXEB & 161  &        9 &   1&Optimized &        3.3 &       13.6 &         6.6 &          3.8 &         5.4 &         6.6 &         7.6 &          10.7 \\
CZXEB & 241  &       11 &   1&Optimized &        2.9 &       15.7 &         7.2 &          4.0 &         5.7 &         6.9 &         8.3 &          11.9 \\
CZXEB & 337  &       13 &   1&Optimized &        3.0 &       74.6 &         7.6 &          3.8 &         5.6 &         6.9 &         8.3 &          11.5 \\
CZXEB & 449  &       15 &   1&Optimized &        2.7 &      191.0 &         8.2 &          3.8 &         5.8 &         6.9 &         8.3 &          12.1 \\
CZXEB & 577  &       17 &   1&Optimized &        3.0 &       53.5 &         7.4 &          4.0 &         5.9 &         7.2 &         8.4 &          11.2 \\
CZXEB & 721  &       19 &   1&Optimized &        3.0 &       33.8 &         7.4 &          4.1 &         6.0 &         7.1 &         8.4 &          11.9 \\
CZXEB & 881  &       21 &   1&Optimized &        2.7 &       66.3 &         7.3 &          4.0 &         5.8 &         7.0 &         8.2 &          11.2 \\
CZXEB & 1057 &       23 &   1&Optimized &        2.9 &       50.9 &         7.4 &          4.2 &         6.0 &         7.2 &         8.4 &          11.8 \\
\hline
CZXEB & 1057 &       23 &  1&Stitched ($R=4$) &        2.6 &       50.4 &         6.5 &          3.4 &         5.2 &         6.3 &         7.5 &          10.6 \\
\hline
CZXEB & 17   &        3 &    1&Baseline &        3.4 &       44.1 &        16.0 &          5.5 &         9.1 &        11.6 &        17.4 &          40.3 \\
CZXEB & 49   &        5 &    1&Baseline &        4.1 &       87.3 &        19.3 &          6.7 &        10.5 &        15.7 &        21.1 &          61.6 \\
CZXEB & 97   &        7 &    1&Baseline &        3.7 &      135.7 &        23.0 &          6.0 &        10.4 &        13.8 &        23.0 &         108.4 \\
CZXEB & 161  &        9 &    1&Baseline &        4.0 &      252.1 &        22.0 &          5.9 &        11.0 &        14.3 &        21.8 &          89.9 \\
CZXEB & 241  &       11 &    1&Baseline &        5.7 &      934.7 &        30.6 &          7.3 &        11.6 &        15.5 &        22.7 &         108.6 \\
CZXEB & 337  &       13 &    1&Baseline &        3.6 &      341.3 &        21.5 &          6.6 &        11.4 &        14.9 &        21.0 &          78.7 \\
CZXEB & 449  &       15 &    1&Baseline &        4.7 &      348.0 &        23.6 &          6.8 &        11.1 &        14.4 &        19.9 &         113.9 \\
CZXEB & 577  &       17 &    1&Baseline &        3.8 &      448.0 &        22.9 &          6.5 &        10.9 &        14.6 &        20.7 &         101.4 \\
CZXEB & 721  &       19 &    1&Baseline &        3.8 &      381.3 &        23.6 &          6.3 &        11.1 &        14.5 &        20.7 &         109.9 \\
CZXEB & 881  &       21 &    1&Baseline &        4.2 &      543.9 &        26.0 &          6.3 &        11.1 &        14.8 &        21.0 &         127.3 \\
CZXEB & 1057 &       23 &    1&Baseline &        3.7 &      888.3 &        23.6 &          6.8 &        11.3 &        14.8 &        20.4 &         115.4 \\
\hline
\hline
SQRB & - & - & - &       Outlier& - &    - &     - &      - &     - &     - &     - &       - \\ 
\hline
SQRB & 17   &        3 &   1&Optimized &      0.53 &      0.96 &       0.72 &        0.56 &       0.63 &       0.71 &       0.83 &         0.89 \\
SQRB & 49   &        5 &   1&Optimized &      0.53 &      2.11 &       0.78 &        0.54 &       0.63 &       0.73 &       0.86 &         1.10 \\
\hline
SQRB & 49 & 5 & - &Crossover &      0.33 &      1.47 &       0.73 &        0.35 &       0.58 &       0.72 &       0.84 &         1.10 \\
\hline
SQRB & 97   &        7 &   1&Optimized &      0.52 &     13.11 &       0.89 &        0.53 &       0.65 &       0.73 &       0.83 &         1.26 \\
SQRB & 161  &        9 &   1&Optimized &      0.51 &      2.44 &       0.81 &        0.54 &       0.69 &       0.78 &       0.88 &         1.38 \\
SQRB & 241  &       11 &   1&Optimized &      0.49 &      1.61 &       0.78 &        0.56 &       0.69 &       0.76 &       0.85 &         1.10 \\
SQRB & 337  &       13 &   1&Optimized &      0.49 &     44.00 &       0.98 &        0.55 &       0.70 &       0.78 &       0.88 &         1.30 \\
SQRB & 449  &       15 &   1&Optimized &      0.43 &     74.91 &       1.00 &        0.55 &       0.69 &       0.77 &       0.88 &         1.29 \\
SQRB & 577  &       17 &   1&Optimized &      0.44 &      4.28 &       0.81 &        0.56 &       0.69 &       0.77 &       0.87 &         1.24 \\
SQRB & 721  &       19 &   1&Optimized &      0.47 &      6.52 &       0.84 &        0.56 &       0.70 &       0.80 &       0.90 &         1.24 \\
SQRB & 881  &       21 &   1&Optimized &      0.44 &     35.55 &       0.84 &        0.56 &       0.70 &       0.79 &       0.87 &         1.19 \\
SQRB & 1057 &       23 &   1&Optimized &      0.45 &      5.58 &       0.83 &        0.56 &       0.70 &       0.79 &       0.90 &         1.23 \\
\hline
SQRB & 1057 &       23 &  1&Stitched ($R=4$) &      0.43 &      5.58 &       0.82 &        0.57 &       0.71 &       0.79 &       0.88 &         1.16 \\
\hline
SQRB & 17   &        3 &    1&Baseline &      0.53 &     10.42 &       1.71 &        0.62 &       0.89 &       1.23 &       1.38 &         2.37 \\
SQRB & 49   &        5 &    1&Baseline &      0.63 &     13.47 &       1.61 &        0.72 &       0.87 &       1.03 &       1.41 &         5.85 \\
SQRB & 97   &        7 &    1&Baseline &      0.52 &     78.62 &       2.80 &        0.57 &       0.91 &       1.11 &       1.44 &        10.47 \\
SQRB & 161  &        9 &    1&Baseline &      0.51 &     78.78 &       2.45 &        0.61 &       0.89 &       1.08 &       1.43 &         6.94 \\
SQRB & 241  &       11 &    1&Baseline &      0.52 &    879.29 &       6.06 &        0.61 &       0.92 &       1.21 &       1.61 &        15.20 \\
SQRB & 337  &       13 &    1&Baseline &      0.46 &     80.89 &       2.41 &        0.58 &       0.95 &       1.28 &       1.67 &         8.21 \\
SQRB & 449  &       15 &    1&Baseline &      0.46 &     80.73 &       2.34 &        0.63 &       0.91 &       1.26 &       1.69 &        10.71 \\
SQRB & 577  &       17 &    1&Baseline &      0.44 &    119.90 &       2.63 &        0.60 &       0.91 &       1.17 &       1.56 &         9.10 \\
SQRB & 721  &       19 &    1&Baseline &      0.48 &    348.00 &       3.10 &        0.63 &       0.93 &       1.22 &       1.68 &        21.58 \\
SQRB & 881  &       21 &    1&Baseline &      0.49 &    393.14 &       4.01 &        0.62 &       0.96 &       1.24 &       1.70 &        15.74 \\
SQRB & 1057 &       23 &    1&Baseline &      0.45 &    199.55 &       3.02 &        0.62 &       0.93 &       1.21 &       1.62 &        13.30 \\
\hline
\end{tabular}
\label{tab:si_bench_scaling_sim}
\end{table*}

\begin{table*}
\centering
\caption{Definitions for symbols used in the main text and supplemental information.}
\begin{tabular}{c|l}
\hline
Symbol & Definition \\
\hline
$S$ & Snake's scope parameter\\
\hline
$q_i$ & Qubit with index $i$ \\
SQ$_i$ & Single-qubit gate executed by $q_i$ \\
CZ$_{ij}$ & Controlled-Z gate executed by $q_i$ and $q_j$\\
$N$ & Number of qubits in a configuration \\
$d$ & Distance of a surface-code logical qubit\\
\hline
$f_i$ & Idle frequency for $q_i$ \\
$f_{ij}$ & Interaction frequency for $q_i$ and $q_j$ \\
$F$ & Set of $f_i$ and $f_{ij}$ corresponding to a frequency configuration\\
$F^*$ & Set of optimized gate frequencies corresponding to $F$\\
$F_g$ & Subset of $F$ that are relevant to estimating the error of gate $g$\\
$F_{g,m}$ & Subset of $F_g$ that are relevant to estimating the error of gate for physical error mechanism $m$\\
$F_S$ & Subset of $F$ selected for optimization by Snake based on scope $S$\\
$k$ & Approximate number of frequency options per gate\\
\hline
$e_i$ & Error per gate for SQ$_i$, measured by SQRB and reported as an average error\\
$e_{ij}$ & Error per gate for CZ$_{ij}$, which can be inferred from SQRB and CZXEB benchmarks\\
$e_{c,ij}$ & Error per cycle for SQ$_i$, SQ$_j$, and CZ$_{ij}$, measured by CZXEB and reported as an average error\\
$e_c$ & Distribution of $e_{c,ij}$ corresponding to some configuration(s)\\
\hline
$N_{sat}$ & Qubit saturation constant in the heuristic saturation scaling model\\
$e_{sat}$ & Saturated error in the heuristic saturation scaling model\\
$e_{scale}$ & Scaling penalty in the heuristic saturation scaling model\\
\hline 
$D$ & Set of characterization data taken by the control system\\
$T_1$ & Energy-relaxation time\\
$T_\phi$ & Dephasing time as inferred from the flux sensitivity $\frac{df}{d\phi}$\\
$\chi$ & Stray coupling parameters between parasitically coupled gates\\
$\delta$ & Frequency-pulse distortion parameters\\
$\mathcal{F}$ & Frequency trajectory used to implement a CZ gate\\
\hline
$A$ & Quantum circuit(s) for the target quantum algorithm(s)\\
$g$ & Some gate in $A$\\
$E$ & Algorithm error estimator for $A$\\
$E_g$ & Gate error estimator for gate $g$\\
$E_S$ & Snake estimator that generally comprises some terms in $E$\\
$M$ & Set of physical error mechanisms $m$\\
$m$ & One physical error mechanism\\
$\epsilon_{g,m}$ & Algorithm-independent error component for gate $g$ and physical error mechanism $m$\\
$w_{g,m}$ & Algorithm-dependent weight corresponding to $\epsilon_{g,m}$\\
\hline
$P$ & Probability density, used in the context of simulated processors \\
$\mathcal{N}$ & Processor size and connectivity, used in the context of simulated processors \\
$\mathcal{P}$ & Processor architectural parameters, used in the context of simulated processors \\
\hline
$\text{argmin}_x(f(x))$ & Minimizer for some function $f(x)$ with respect to its argument(s) $x$\\
$|.|$ & Cardinality of the set $.$\\
$\langle . \rangle$ & Expectation value of .\\
$\sigma$ & Standard deviation\\
\hline
\end{tabular}
\label{tab:definitions}
\end{table*}

\clearpage
\bibliography{main.bib}